\documentclass[a4paper,11pt]{article}
\pdfoutput=1 

\usepackage{jcappub} 

\usepackage[T1]{fontenc} 
\usepackage{xcolor}
\usepackage{multirow}
\usepackage{soul}

\title{Cosmological Constraints on sub-horizon scales modified gravity theories with MGCLASS II}

\author[a]{Z. Sakr}
\author[b,c]{M. Martinelli}

\affiliation[a]{Faculty of Sciences, Universit\'e St Joseph; Beirut, Lebanon}
\affiliation[b]{INAF - Osservatorio Astronomico di Roma, via Frascati 33, 00040 Monteporzio Catone (Roma), Italy}
\affiliation[c]{Instituto de F\'isica T\'eorica UAM-CSIC, Campus de Cantoblanco, 28049 Madrid, Spain}

\emailAdd{ziad.sakr@net.usj.edu.lb}
\emailAdd{matteo.martinelli@inaf.it}

\abstract{In this paper we introduce a new public Einstein-Boltzmann solver, \texttt{MGCLASS II}, built as a modification of the publicly available \texttt{CLASS} code, that allows to obtain cosmological observables in Modified Gravity theories. It implements several commonly used parameterizations of deviations from General Relativity, computing their impact on the growth of structure as well as on the background evolution of the Universe, together with a subset of available alternative theories, still not completely ruled out by observations. \texttt{MGCLASS II} is built in such a way to be compatible with parameter estimation codes such as \texttt{MontePython} and \texttt{Cobaya}. We exploit this possibility to constrain the parameterizations used by the Planck collaboration, in order to validate the predictions of this new code, and a newly implemented parameterization (z\_flex) which has different features. For the former we find good agreement with the results existing in the literature, while we present original constraints on the parameters of the latter, finding no significant deviation from the standard cosmological model, $\Lambda$CDM.}

\begin{document}
\maketitle
\flushbottom

\section{Introduction}\label{sec:intro}

Since General Relativity (GR) was proposed in 1915 and applied to the evolution of the universe in 1917, the progress in observational techniques has allowed to extensively test this theory both within the Solar System and at cosmological scales. Einstein's theory was able to pass all these tests; however, the observations of the luminosity-redshift evolution from Supernovae Ia made in 1998 \cite{1998AJ....116.1009R,1999ApJ...517..565P}, followed later by other measurements from other cosmological probes, strongly supports an accelerated expanding Universe, which cannot be explained within GR without adding a cosmological constant ($\Lambda$) in the Einstein field equations. The simplest explanation for this constant, by means of negative pressure from vacuum energy, suffers from a significant discrepancy between its observed value and the one calculated in the context of quantum field theory \citep{Weinberg:1988cp}. Moreover the initial value of the dark energy density needs fine-tuning to reach nowadays the value constrained from different cosmological observables, like CMB temperature and polarization angular correlations \cite{Aghanim:2018eyx}, baryonic acoustic oscillation imprints on galaxy correlations (BAO) \cite{Alam:2020sor} or the aforementioned luminosity-redshift diagram determined from supernovae observations \cite{Scolnic:2017caz}.

Modifications of gravity at cosmological scales were proposed to remedy for these issues, with theories alternative to GR developed in such a way to account for the late time accelerated phase of the expansion of the Universe, with the requirement that these reduce to GR at Solar System scales, where the theory of gravity is tightly constrained to be close to GR. The investigation of these possible alternatives can follow different approaches; on the one hand, one can develop a theory of gravity by modifying the Einstein-Hilbert action, thus breaking the premises of the Lovelock theorem; this states that GR is the only gravitational theory with second order or less differential equations that can be constructed in a four-dimensional Riemannian space from an action principle involving the metric tensor and its derivatives only \cite{Lovelock:1971yv}. The advantage of this approach is to have alternative models that stem from first principles, but has the disadvantage to be computationally expensive since each model must be separately constrained and its performance compared with that of GR. Alternatively, one could consider an Effective Field Theory approach (EFT) which allows to define a general action for a broad class of modified gravity theories, and therefore can be used to derive a set of equations that encodes the behaviour of several alternative models (see \cite{Frusciante:2019xia} and references therein). Finally, other investigations rely instead on a purely phenomenological approach; in this case one modifies the main equations ruling the background expansion and perturbations evolution, introducing free functions that encode possible deviations from the GR$+\Lambda$ paradigm. A common example is the parameterisation of the evolution of the dark energy (DE) equation of state using the first two terms of a Taylor expansion \cite{2001IJMPD..10..213C,2003PhRvL..90i1301L}, while for the perturbations, two functions that alter the relations between the Newtonian and Weyl potential and the matter density contrast can be introduced \cite{Bardeen:1980kt,Mukhanov:1990me,Zhang:2007nk,Malik:2008im,Amendola:2007rr,Pogosian:2010tj}. Such an approach has the advantage of being able to test deviations from GR without the need to assume a theoretical model, but the price to pay is the inability to connect such departures to a physical model, or even to know if the deviations considered are physically viable.

Recently, attempts to combine the latter two approaches have been made, with the purpose of preserving as much as possible the model independence of the phenomenological approach, but imposing conditions on the parameterized departures through the EFT in order to ensure their physical viability \cite{Peirone:2017lgi,Gerardi:2019obr,Traykova:2021hbr,Pogosian:2021mcs,Raveri:2021dbu}.

Independently of the chosen approach, in order to perform cosmological analyses and parameter estimation, it is necessary to obtain predictions for cosmological observables in these modified gravity models; this is usually done modifying the public softwares for cosmological predictions \texttt{CAMB} \cite{Lewis:1999bs,2012JCAP...04..027H} and \texttt{CLASS} \cite{2011JCAP...07..034B}. Several modifications of these softwares are publicly available, with e.g. \texttt{EFTCAMB} \cite{Hu:2013twa,Raveri:2014cka} and \texttt{HiCLASS} \cite{Zumalacarregui:2016pph,Bellini:2019syt} implementing the EFT approach, or \texttt{MGCAMB} \cite{Zucca:2019xhg}, which instead relies on the phenomenological approach. We also note other modified gravity patches: \texttt{EFCLASS} \cite{Arjona:2018jhh} that follows an effective fluid approach, \texttt{CLASSEOSFR} \cite{Battye:2017ysh} with an equation of state approach and \texttt{QSACLASS} \cite{Pace:2020qpj} which implements a quasi-static approximation to Hordenski models. 

In this paper, we will focus on the phenomenological approach, but instead of relying on the existing \texttt{MGCAMB} code, we use our own modification of \texttt{CLASS}, \texttt{MGCLASS II}. This new code is built, in a coherent and consistent way within the Boltzmann solver, starting from a first modification of \texttt{CLASS} that implements some phenomenological parameterizations of modified gravity, presented in \cite{Baker:2015bva}. We significantly extend the range of models and parameterizations available in this new version, also including many that are present in \texttt{MGCAMB}, due to their usefulness and importance in the landscape of the Modified Gravity models, but also in order to facilitate the comparison between the two codes.

\texttt{MGCLASS II} is made publicly available\footnote{\url{https://gitlab.com/zizgitlab/mgclass--ii}}, and the usefulness of it, other than the possibility to choose different parameterizations, will be to cross-validate the predictions of existing codes with those of a completely independent software, and to interface with \texttt{CLASS} based data analysis softwares.

The paper is structure as follows. In \autoref{sec:MGgeneralities} we review the main equations used in the phenomenological approach to modified gravity and we show how, in particular cases, the common way to parameterize deviations from the standard perturbations evolution can be used to infer how alternative models would impact also the background expansion. In \autoref{sec:parameterizations} we show a pair of commonly used parameterizations, for which results in the literature already exist, and we introduce a new one, which has different features and can provide complementary information. This exercise has the double purpose to obtain novel results and, at the same time, validate the modifications of the standard equations on which \texttt{MGCLASS II} is based. Such modifications to \texttt{CLASS} are presented in \autoref{sec:code}, where we also highlight the main features of the code. In \autoref{sec:analysis} we describe our pipeline for parameter estimation, we present the results obtained on the considered scenarios, and we draw our conclusions in \autoref{sec:conclusions}. In addition to this, in \autoref{sec:appendix} we comment on the further differences with respect to some specific codes that are not discussed in the main body of the paper.

\section{Phenomenological paramerizations of modified gravity in the quasi static assumption}\label{sec:MGgeneralities}

In this section we summarize the impact of Modified Gravity theories on the background expansion of the Universe and on the evolution of cosmological perturbations. We focus here on a broad class of such theories, introducing an additional scalar degree of freedom $\phi$ in the action.

Starting from a general 4-dimensional action
\begin{equation}\label{eq:general_action}
S=\int d^4x \sqrt{-g}\left(\frac12f(R,\phi,X)+\mathcal{L}_m\right),
\end{equation}
with $X=-g^{\mu\nu}\partial_\mu \phi \partial_\nu \phi$ the kinetic term of the additional scalar field and $f$ a generic function, one can specify to a class of theories where the gravity Lagrangian is given by
\begin{equation}
\mathcal{L}=\frac{F(\phi)}{2}R+X-U(\phi)\, , \label{eq:SCTaction}
\end{equation}
where $U(\phi)$ is the potential of the scalar field and $F(\phi)$ a function determining its coupling to gravity. Such a class of theories encompasses popular models such as $f(R)$ \cite{Buchdahl:1983zz} and scalar-tensor theories \cite{Horndeski:1974wa,Kobayashi:2019hrl}

While it is possible to specify to a particular model by defining the terms $U(\phi)$ and $F(\phi)$, it is often preferred to explore possible deviations from GR in a parametric way. It can be shown \cite{Mukhanov:1990me,Zhang:2007nk,Malik:2008im,Amendola:2007rr} that this class of modified gravity theories impacts the evolution of the Universe by modifying the Poisson and anisotropic stress equations, which determine the evolution of cosmological perturbations, i.e.

\begin{align}
 -k^2\,\Psi(a,\vec{k}) &= \frac{4\pi\,G}{c^4}\,a^2\,\bar\rho(a)\,\Delta(a,\vec{k})\times\mu(a,k), \label{eq:mu}\\ 
 \Phi(a,\vec{k}) &= \Psi(a,\vec{k})\times\eta(a,k),  
  \label{eq:eta}\\
 -k^2\,\left[\Phi(a,\vec{k})+\Psi(a,\vec{k})\right] &= \frac{8\pi\,G}{c^4}\,a^2\,\bar\rho(a)\,\Delta(a,\vec{k})\times\Sigma(a,k), \label{eq:sigma}
\end{align}
where $\bar\rho\Delta=\bar\rho\delta+3(aH/k)(\bar\rho+\bar p)v$ is the comoving density perturbation of $\delta=(\rho-\bar{\rho})/\bar{\rho}$, and $\rho$, $p$ and $v$ are, respectively, the density, pressure and velocity with the bar denoting mean quantities. Note that in the equations above we neglect the anisotropic stress due to relativistic species for simplicity; however, even though this effect is negligible at late times, the equations used in our code account for such effect in the evolution of the potentials $\Phi$ and $\Psi$. These are the Bardeen potentials entering the perturbed FLRW metric, which in Newtonian gauge is
\begin{equation}
ds^2=a^2\left[-(1+2\Psi)d\tau^2+(1-2\Phi)d\vec{x}^2\right]\,.
\end{equation}
The free functions $\mu(a,k)$, $\eta(a,k)$ and $\Sigma(a,k)$ encode the possible deviations from GR, which is recovered when all these three are constant and equal to unity. Of the three functions, only two are independent, as the system of \autoref{eq:mu}, \autoref{eq:eta} and \autoref{eq:sigma} reduces to
\begin{equation}
    \Sigma(a,k) = \frac{\mu(a,k)}{2}\left(1+\eta(a,k)\right)\, .
\end{equation}

The deviations from GR encoded in these functions can be written in terms of the additional degree of freedom $\phi$ entering \autoref{eq:SCTaction}. However, this requires to assume the quasi-static approximation (QSA) and to restrict the analysis to times way beyond matter domination where MG affects sub-horizon scales. Within such assumptions, the functions can be rewritten as  \cite{Tsujikawa:2007gd}

\begin{align}
\mu(a,k) &= \frac{1}{8\pi F}
\frac{f_{,X}+4\left(f_{,X}\frac{k^2}{a^2}\frac{F_{,R}}{F}+
\frac{F_{,\phi}^2}{F}\right)}
{f_{,X}+3\left(f_{,X}\frac{k^2}{a^2}\frac{F_{,R}}{F}+
\frac{F_{,\phi}^2}{F}\right)}\,, \label{eq:musub}\\ 
 \eta(a,k) &=\frac{2f_{,X}\frac{k^2}{a^2}\frac{F_{,R}}{F}
+\frac{2F_{,\phi}^2}{F}}
{f_{,X} \left(1+\frac{2k^2}{a^2}\frac{F_{,R}}{F}\right)+
\frac{2F_{,\phi}^2}{F}}\, , \label{eq:etasub}
\end{align}

where, following \autoref{eq:SCTaction}, $F=F(\phi)=\partial_R f(R,\phi,X)$, $F_{,\phi}=\partial_\phi F(\phi)$ and $f_{,X}=\partial_Xf(R,\phi,X)$. It is therefore clear how one can express the phenomenological MG functions in a particular model by specifying the functions $F(\phi)$ and $U(\phi)$ entering \autoref{eq:SCTaction}.

While most of the parametric investigations of modified gravity assume a background indistinguishable from that of $\Lambda$CDM (see e.g.\cite{Ade:2015rim,Aghanim:2018eyx}), these extended theories impact in principle also the expansion of the Universe.  Considering such an effect would allow to improve the constraints that are achievable on such theories, by combining the information brought by perturbations and background observables.

Using the homogeneous FLRW metric in the action \autoref{eq:SCTaction} and assuming a homogeneous scalar field without interactions with dark matter and a perfect fluid background with no clustering, we find the dynamical equations of the system as \cite{Tsujikawa:2007gd}
\begin{align}
3F H^2 &=  \rho_m +{1\over 2} \dot\phi^2 - 3 H \dot F + U\,\label{eq:H_eff_1}\, ,\\ 
-2F \dot{H} &= (\rho_{\Lambda}+p_{\Lambda})+ \ddot{F} - H\dot{F} + \rho_{\rm tot}\, ,
\label{eq:H_eff_2}
\end{align}
 where the dot represents the derivative with respect to cosmic time $t$ and $\rho_{\rm tot}$ is the total energy density.

Since a $\Lambda$CDM background is compatible with cosmological data, we choose here $U=2\Lambda$ where $\Lambda$ is the cosmological constant; this implies that the contribution coming from modified gravity introduces deviations from the standard model expansion, without replacing the cosmological constant. In the QSA we can also consider $\dot \phi \sim 0$. In order to solve the equation for $H$ and obtain the expansion history of the Universe, we need to determine the function $F$ and its derivative. By substituting our choice for $U$ in \autoref{eq:musub} and \autoref{eq:etasub}, the MG functions become \cite{Nesseris:2017vor}
 \begin{align}
\mu(a,k) &= \frac{1}{F(\phi)}\frac{F(\phi)+2F_{,\phi}^2}{F(\phi)+\frac32F_{,\phi}^2}\, ,\label{eq:mueff}\\
\eta(a,k) &= \frac{F_{,\phi}^2}{F(\phi)+2F_{,\phi}^2}\, ,\label{eq:etaeff}
\end{align}
which can be solved for $F$ and $\dot F$ finding 
\begin{align} 
 F &= \frac{2}{\mu + \mu\eta}\, ,\label{eq:eff_1}\\ 
 \dot F &= - \frac{2(\dot \mu (1+\eta)+ \mu \dot\eta)}{(\mu + \mu\eta)^2}\, ,\label{eq:eff_2}\\
 \ddot F &= 4 \frac{(\dot \mu(1+\eta) + \mu \dot \eta)^2}{(\mu + \mu\eta)^3} - 2 \, \frac{2 \,\dot \mu \dot \eta + \dddot \mu (1+\eta)+\mu \ddot \eta}{(\mu + \mu\eta)^2}\, \label{eq:eff_3}\, .
\end{align}

This means that specifying the redshift trends of $\mu$ and $\eta$, either by choosing a model or a parameterization, will allow to obtain the modified background expansion of the Universe through \autoref{eq:H_eff_1} and \autoref{eq:H_eff_2}.

\section{Parameterized modified gravity}\label{sec:parameterizations}

The discussion of \autoref{sec:MGgeneralities} has shown how the MG functions $\mu$, $\eta$ and $\Sigma$ can encode the phenomenology of a significantly large class of alternative theories of gravity. If one specifies to a given model, it is possible to compute these functions and obtain its impact on cosmological observables. However, mapping a model in the MG functions requires some assumptions to be made. If one wants to avoid these and, moreover, to remain general without specifying a particular model it is possible to simply investigate deviations from the GR limit $\mu=\eta=\Sigma=1$ by parameterizing these functions \cite{Daniel:2010ky,Zhao:2010dz,Simpson:2012ra} or reconstructing them through a binned approach \cite{Pogosian:2021mcs,Raveri:2021dbu,Hojjati:2011xd,Kennedy:2018gtx,Raveri:2019mxg}.

One of the most commonly used parameterizations expresses the MG functions in a way that is typical of theories encompassed in the Horndeski class \cite{Zhao:2008bn}, i.e.
\begin{eqnarray}
\mu(k,z) = 1 + f_1(z)\frac{1+c_1\left(\frac{\mathcal{H}}{k^2}\right)}{1+\left(\frac{\mathcal{H}}{k^2}\right)}\, , \label{eq:plkmu}\\
\eta(k,z) = 1 + f_2(z)\frac{1+c_2\left(\frac{\mathcal{H}}{k^2}\right)}{1+\left(\frac{\mathcal{H}}{k^2}\right)}\, ,\label{eq:plketa}
\end{eqnarray}
where $\mathcal{H} = H/(1+z)$ is the comoving Hubble parameter.
Such a parameterization has been widely used by observational collaborations to constrain possible deviations from GR, e.g. by Planck using CMB data \cite{Ade:2015rim,Aghanim:2018eyx} and DES using Large Scale Structures observations \cite{Abbott:2018xao}. In \autoref{eq:plkmu} and \autoref{eq:plketa}, the functions $f_i(z)$ regulate the amplitude in redshift of such deviations, while the $c_i$ parameters affect their scale dependence.

Given that we expect the MG functions to reduce to their GR limit at early times, as the modifications to the theory of gravity should be relevant only at small redshifts, it is common to assume that the amplitude of the modifications scales with the dark energy density $\Omega_{\rm DE}(z)$ \cite{Ade:2015rim,Aghanim:2018eyx}, i.e.
\begin{equation}\label{eq:plklate}
    f_i(z)=E_{ii}\Omega_{\rm DE}(z)\, .
\end{equation}
One can however allow more freedom to this parametric expression using assumptions that do not automatically return to GR at early times, such as \cite{Ade:2015rim}
\begin{equation}\label{eq:plk_early}
    f_i(z) = E_{i1} + E_{i2} \frac{z}{1+z}\,.
\end{equation}
We will refer to these two possible parameterizations as "Planck late" and "Planck early" respectively.

When using these parameterizations throughout the rest of the paper, we set ourselves in the scale independent limit, where $c_1=c_2=1$. This is motivated by the fact that we will mainly use CMB data to constrain such parameterizations, and it has been found that these do not allow to constrain the scale dependence of these MG models \cite{Ade:2015rim}.

\subsection{z\_flex parameterization}\label{sec:zxpans}

The two parameterizations shown above are distinguished by their behaviour at high redshift, with "Planck late" returning to GR as $\Omega_{\rm DE}(z)$ approaches zero, and "Planck early" allowing for more freedom for the redshift trend of the MG functions, disentangling it from the background evolution. However the high redshift limit of the latter parameterization is always non-vanishing, except for the case where $E_{i1}+E_{i2}=0$. Therefore, it does not generally reduce to GR at early times where one would expect all modifications to vanish, and this feature could prevent this parameterization to catch deviations from GR at late times, hence the label 'early' for the model.

We thus consider also another parameterization, extending one introduced in \cite{Nesseris:2017vor}, which automatically returns to GR at high redshifts without however the necessity of connecting it to the background evolution of $\Omega_{\rm DE}(z)$:
\begin{align}\label{eq:zxpanspar}
 \mu(a)  &= 1+g_{\mu}(1-a)^n - g_{\mu}(1-a)^{2n}\, ,\\   
 \eta(a)  &= 1+g_{\eta}(1-a)^n - g_{\eta}(1-a)^{2n} \, , 
\end{align}
where $g_\mu$, $g_\eta$ and $n$ are free parameters.

One can notice how for $a\rightarrow0$, the two MG functions automatically tend to a unit value, with the smoothness of the transition regulated by the parameter $n$, and $g_i$ the amplitude of the deviations. In addition to this, this parameterization also gives no deviations from GR at present time. These two features imply that a MG investigation using this parameterization satisfies both Solar System and Big Bang Nucleosynthesis (BBN) constraints, thus only allowing the gravitational mechanism to depart GR only at intermediate redshifts.

Notice that Solar System tests can also be satisfied by the "Planck late" and "Planck early" parameterizations, by assuming that some screening mechanism effectively reduces the MG theory to GR in our local environment \cite{Khoury:2010xi,Brax:2013ida,Falck:2014jwa}.

\section{MGCLASS II code description}\label{sec:code}

\texttt{MGCLASS} II is an extension to the cosmological and Boltzmann solver \texttt{CLASS} allowing to incorporate the linear effects of 
different classes of modified gravity theories, mainly within the quasi-static assumption and on their sub horizon range. The purpose of this code is to obtain theoretical predictions for cosmological observables in alternative theories of gravity, either using specific theoretical models or through parameterized deviations from GR. \texttt{MGCLASS} II is an evolution of a first version by \citep{Baker:2015bva} in which modified gravity effects are modeled by introducing two functions that parameterize MG effects through an alteration of the relations between the matter density contrast and the Newtonian potential, as in \autoref{eq:mu}, and between the Newtonian potential and the curvature perturbation, as in \autoref{eq:eta}. This new version significantly expands the number of parameterization that can be selected by the user, and also implements several specific models of alternative theories of gravity. 
In addition to this, the code is brought up to speed with a more recent version of \texttt{CLASS} (2.9) and implements different conditions that can be imposed to the selected deviations from GR (see \autoref{sec:features}). Finally, \texttt{MGCLASS II} also allows to account for the effects of deviations from GR on the background expansion of the Universe, following the approach described in \autoref{sec:MGgeneralities}.

\texttt{MGCLASS II} adds new features to the \texttt{modgrav.c} module which was added to the basic structure of \texttt{CLASS} in \cite{Baker:2015bva}. This module computes the trends in redshift and scale of the three MG functions $\mu(z,k)$, $\eta(z,k)$ and $\Sigma(z,k)$ depending on the model chosen by the user. These functions are then used in the modified equations for perturbations and background evolutions, implemented in \texttt{perturbations.c} and \texttt{background.c} respectively, that we describe below.

The \texttt{MGCLASS II} code is publicly available\footnote{\url{https://gitlab.com/zizgitlab/mgclass--ii}}, and an example jupyter notebook is provided together with the main code, in order to offer a walkthrough for users on how to use the different models and options implemented. As we describe below in more detail, the modified equations for perturbations and background evolution that we implemented are fairly general; this has the advantage to allow users to implement new models or parameterizations in a very simple way, by just adding new possible options for the calculation of $\mu(z,k)$ and $\eta(z,k)$ in \texttt{modgrav.c}, and the corresponding parameters in \texttt{input.c}.

As it is based on \texttt{CLASS}, \texttt{MGCLASS II} is easily interfaced with cosmological data analysis codes that make use of the standard code, e.g. \texttt{MontePython} \cite{2013JCAP...02..001A} and \texttt{Cobaya} \cite{Torrado:2020dgo}, thus allowing to use this code to obtain constraints on modified gravity parameters.

\subsection{Modifications for perturbations evolution}

The main module in \texttt{CLASS} computing predictions for the evolution of cosmological perturbations is \texttt{perturbations.c}. The equations used can be obtained from the line element of the perturbations of the FLRW metric in a general gauge
\begin{align}
ds^2&=a^2(\eta)\Big\{-(1+2A)d\eta^2-2B_i dx^id\eta \nonumber\\
&+\left[(1+2H_L)\delta_{ij}+2H_{T\,ij}\right]dx^idx^j\Big\},
\end{align}
where $A$ is a scalar potential, $B_i$ a vector shift, $H_L$ a scalar perturbation to the spatial
curvature and $H_{T\,ij}$ a trace-free distortion to spatial metric.

Specifying to the Newtonian gauge, we can obtain a system of equations for cosmological perturbations for the spin-0 components:
\begin{align}
\Psi &= A-\frac{{\cal H}}{k}\left(\frac{\dot{H}_T}{k}-B\right)-\frac{1}{k}\left(\frac{\ddot{H}_T}{k}-\dot{B}\right)\, ,\\
\Phi &= -H_L-\frac{1}{3}H_T+\frac{{\cal H}}{k}\left(\frac{\dot{H}_T}{k}-B\right)\, ,\\
D &= \delta+3(1+w)\frac{{\cal H}}{k}(v-B)\label{Ddef}\, ,\\ 
V &= v-\frac{\dot{H}_T}{k}\, ,\\
Y &= D-3(1+w)\left(\frac{{\cal H}}{k}V+\Phi\right)\, ,
\end{align}
which we close with the standard fluid equations for matter perturbations
\begin{align}
\dot{Y}_{\rm M} &= -kV_{\rm M} \label{fld1}\, ,\\
\dot{V}_{\rm M} &= -{\cal H} V_{\rm M}+k\Psi\label{fld2}.
\end{align}

As already mentioned in \autoref{sec:MGgeneralities}, we are neglecting here the contribution of anisotropic stress from relativistic species for the sake of simplicity. In our code however, these effect is included through the evolution equations for $\Phi$ and $\Psi$ implemented in the standard \texttt{CLASS} code. We also note that our modifications take place after a threshold redshift (\texttt{mg\_z\_init}) that can be selected by the user. Such a redshift is usually well below the matter-radiation equality, since we are interested in low redshift modifications, and therefore our modified equations act in a regime where pressureless matter is the dominant clustering matter component, thus leading to a negligible contribution from anisotropic stress. 

By combining these equations with  the modified Poisson equations presented in \autoref{sec:MGgeneralities}, we obtain a modified evolution for the potential $\Psi$

\begin{align}
\label{PhiYrel}
-2k^2\dot\Psi &= 3{\cal H}^2\Omega_{\rm M}\frac{\left(Y_{\rm M}-3\frac{\cal H}{k^2}\dot{Y}_{\rm M}\right)}{\left(1+\frac{9}{2}\frac{{\cal H}^2}{k^2}\Omega_{\rm M}\eta\right)}.
\end{align}

In the same way, we can obtain the modified evolution for the potential $\Phi$, which reads

\begin{align} \label{eq:newPhidot}
\dot{\Phi} =& \left ( 1 + \frac{9}{2}\frac{aH^2}{k^2} \Omega_{\rm M}  \eta \right )^{-1} \left [ \Phi \left ( \frac{\dot{\eta}}{\eta} + \frac{\dot{\tilde{\mu}}}{\tilde{\mu}} - aH \right ) \right. \nonumber \\
& + \left.\frac{9}{2}\frac{a H^2}{k^2} \Omega_{\rm M}  \eta V \left ( \frac{k}{3} + \frac{a H^2 - a\dot{H}}{k}\right ) - \frac{9}{2}\frac{a H^2}{k^2} \Omega_{\rm M}  \eta \Psi \,aH \right],
\end{align}

The slip relation \ref{eq:eta} is then used as the closure equation to obtain $\Psi$ and $\Phi$.

\texttt{MGCLASS II} implements these general equations in \texttt{perturbations.c}. The specific functional form of $\mu(z,k)$ and $\eta(z,k)$ is instead computed in \texttt{modgrav.c}, which is therefore the main module that needs to be modified should an additional model be implemented, together with \texttt{input.c} where the additional parameters need to be defined.

As stated above, \texttt{MGCLASS II} does not replace completely the standard perturbation equations of \texttt{CLASS}; these are used until the user-defined redshift \texttt{mg\_z\_init}, below which the code switches to the modified equations.

As the modified evolution for $\Phi$ and $\Psi$ are obtained in the Newtonian gauge, \texttt{MGCLASS II} can only work in this specific gauge\footnote{As \texttt{CLASS} is able to work also in synchronous gauge, an error handler stops the code should the user select this option.}. Notice that, as \texttt{MGCAMB} works instead in synchronous gauge, the existence of these two codes allows any user to work in their preferred gauge and include their modifications in the most suited code.

\subsection{Modifications for the background expansion}

As described in \autoref{sec:MGgeneralities}, MG models can in principle affect also the background expansion of the Universe. In order to account for this in \texttt{MGCLASS II}, we modify the main background module of \texttt{CLASS}, i.e. \texttt{background.c}. In this module, we implement the modified Friedmann's equations as
\begin{align}
3F H^2 &=  \rho_{tot}  - 3 H \dot F + \rho_{\Lambda}\, , \\
-2F \dot{H} &=(\rho_{\Lambda}+p_{\Lambda})+ \ddot{F} - H \dot{F} + \rho_{tot}\, .
\end{align}
 
The function $F$ and its derivatives can be connected to the MG functions $\mu$ and $\eta$ as
\begin{align}
 F &= \frac{2}{\mu + \mu\eta} \, ,\\
 \dot{F} &= - \frac{2(\dot{\mu} (1+\eta)+ \mu \dot{\eta})}{(\mu + \mu\eta)^2}\, ,\\
 \ddot{F} &= - 4 (\dot{\mu} + \eta\dot{\mu} + \mu\dot{\eta})^2 (\mu + \mu\eta)^{-3}  - 2 (2\dot{\mu}\dot{\eta} + \ddot{\mu} + \eta\ddot{\mu} 
              + \mu\ddot{\eta})
                     (\mu + \mu\eta)^{-2}  
\end{align}
As for the modifications concerning perturbations evolution, these modified equations are used only below the redshift at which the modifications to GR are activated, and they apply for any choice of $\mu(z,k)$ and $\eta(z,k)$, that are instead computed, together with their derivatives, in \texttt{modgrav.c} according to the model chosen by the user.

Although in principle this method could be applied also to any specific model that can be mapped to a scalar field-like modification of GR, when implementing in the code three specific models ( JBD \citep{Brans:1961sx}, nDGP \citep{Dvali:2000hr} and K-mouflage \citep{Babichev:2009ee}) we choose to follow a designer approach, adopted in associated studies, in which a specific functional form to parameterise the variation of the field with redshift was used for each model. We describe more in details their implementation in \autoref{sec:appendix}.

Another feature of the code is that it does not employ a shooting method to match the cosmological parameters in the modified background evolution, but rather applies analytical methods. This allows to lower the computational time necessary for such operations.

\subsection{MG models and main code's features}\label{sec:features}

The possible deviations from GR implemented in \texttt{MGCLASS II} span an extensive range of possibilities; we consider specific alternative gravity theories which are constructed from first principle and whose behaviour is obtained from their Lagrangian formulation \cite{Hu:2007nk,Brax:2013fna,2012JCAP...10..002B}, with some of these complemented with a designer approach that is used to model their impact on the background expansion \cite{Brans:1961sx,Lima:2015xia,Dvali:2000rv,Koyama:2005kd,Babichev:2009ee,Brax:2015lra}; in addition to these we also consider several parameterization inspired from effective field theory approaches \cite{Bellini:2014fua,Amendola:2012ky,Silvestri:2013ne,Dossett:2015nda,Dossett:2011tn}, and others that are instead purely phenomenological \cite{Ade:2015rim,Baker:2015bva,Bertschinger:2008zb,Zhao:2008bn,Hojjati:2011ix,Alam:2015rsa,Nesseris:2017vor,2010PhRvD..81h3534B}; furthermore, we include a consistent implementation of the growth index parameterisation \cite{2005PhRvD..72d3529L,Hojjati:2011ix} which is not limited to the case where the background expansion mimics that of a cosmological constant, but can also be used when the dark energy equation of state is allowed to deviate from $w=-1$. A more detailed description of the new models and parameterizations implemented is presented in \autoref{sec:appendix}. Note that, in \texttt{MGCLASS II}, all the MG models implemented are compatible with a non-vanishing neutrino mass.

\texttt{MGCLASS II} also allows the user to impose direct conditions on the phenomenological parameterizations implemented. These relate together the three MG functions $\mu(z,k)$, $\eta(z,k)$ and $\Sigma(z,k)$ or manually set their values depending on the selected option, thus setting different conditions on the modifications of the perturbations potentials \citep{Linder:2020xza}, namely: \\

\begin{itemize}
    \item \texttt{no run}: setting the Newtonian potential modified function to its GR value ($\mu(z,k)=1$);
    \item \texttt{no slip}: setting the ratio between the two potentials to be one, thus effectively enforcing $\eta(z,k)=1$;
    \item \texttt{no lens}: forces the code to work with a GR-like gravitational lensing ($\Sigma(z,k)=1$), thus imposing a condition $\eta(k,z) = -1+2/\mu(k,z)$ that introduces a relation between otherwise free functions.
\end{itemize}

\begin{figure}[!h]
	\centering
	\begin{tabular}{cc}
	\includegraphics[width=0.45\columnwidth]{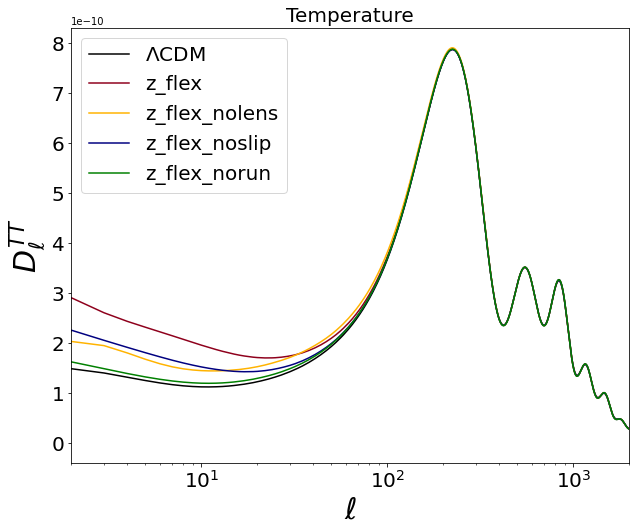} &
	\includegraphics[width=0.45\columnwidth]{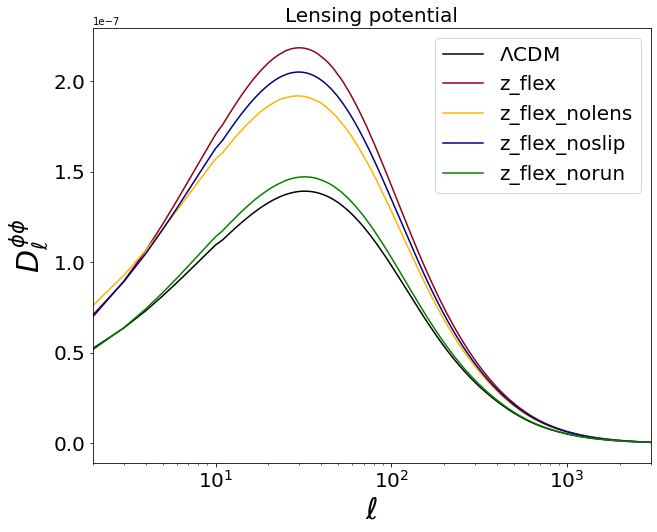}
	\end{tabular}
	\caption{Comparison of the impact of the different phenomenological assumptions available in \texttt{MGCLASS II} for the z\_flex parameterization. The left panel shows the temperature power spectra, while the right panel shows the lensing potential power spectra. For this plot, the cosmological parameters are fixed to the mean values of Planck 2018 \cite{Aghanim:2018eyx}, while the z\_flex are set to be $\{g_\mu,g_\eta,n\}=\{1.0,1.0,1.0\}$}
	\label{fig:pheno_options} \end{figure}

In \autoref{fig:pheno_options} we show the impact of these different options when applied to the z\_flex parameterization of \autoref{sec:zxpans}. The left panel shows the impact on the CMB temperature power spectrum, while the right panel highlights the effect of the conditions on the CMB lensing potential power spectrum.

 \subsection{Validation with \texttt{MGCAMB}}\label{sec:validation}
 
\begin{figure}[!h]
	\centering
	\begin{tabular}{cc}
	\includegraphics[width=0.45\columnwidth]{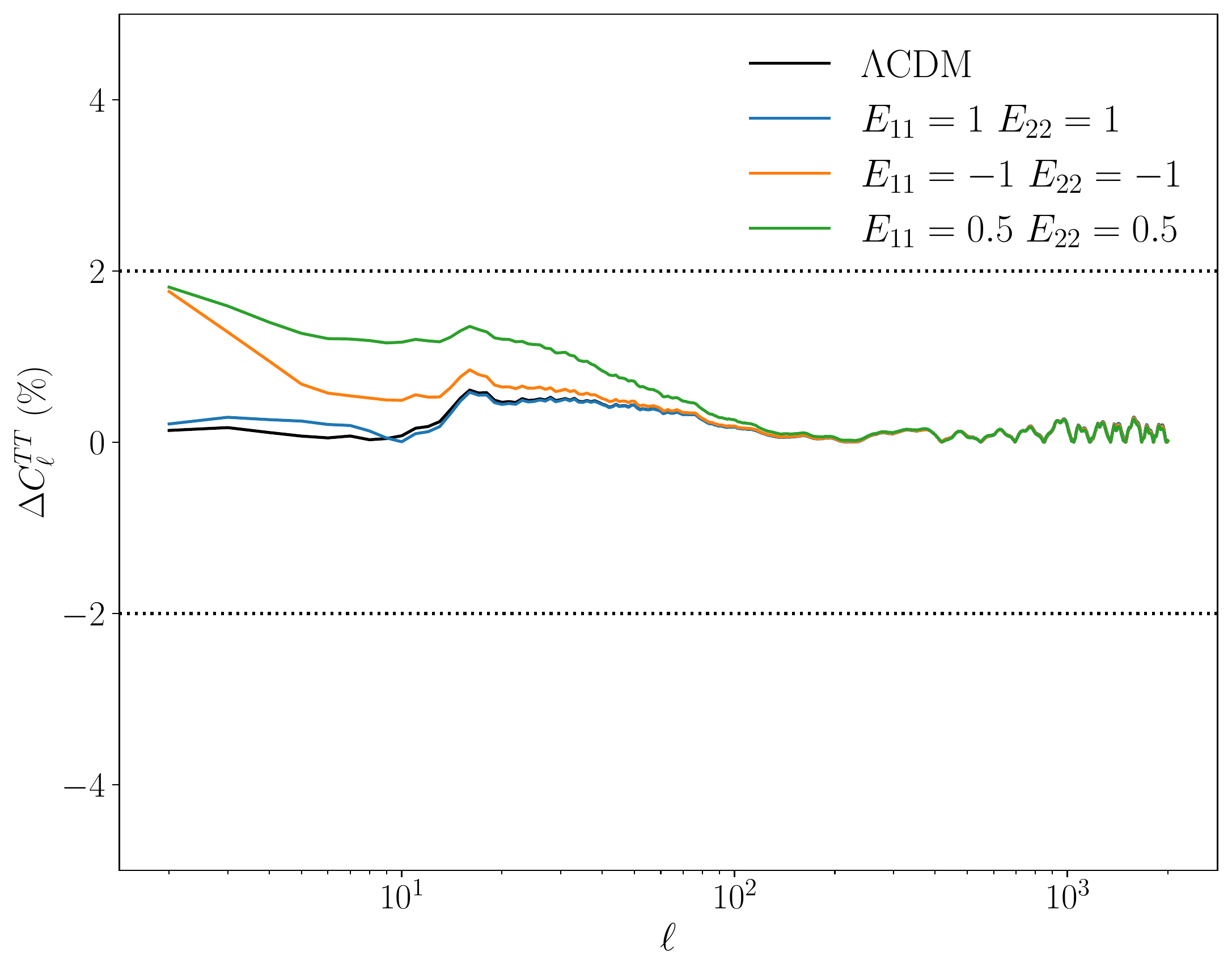} &
	\includegraphics[width=0.45\columnwidth]{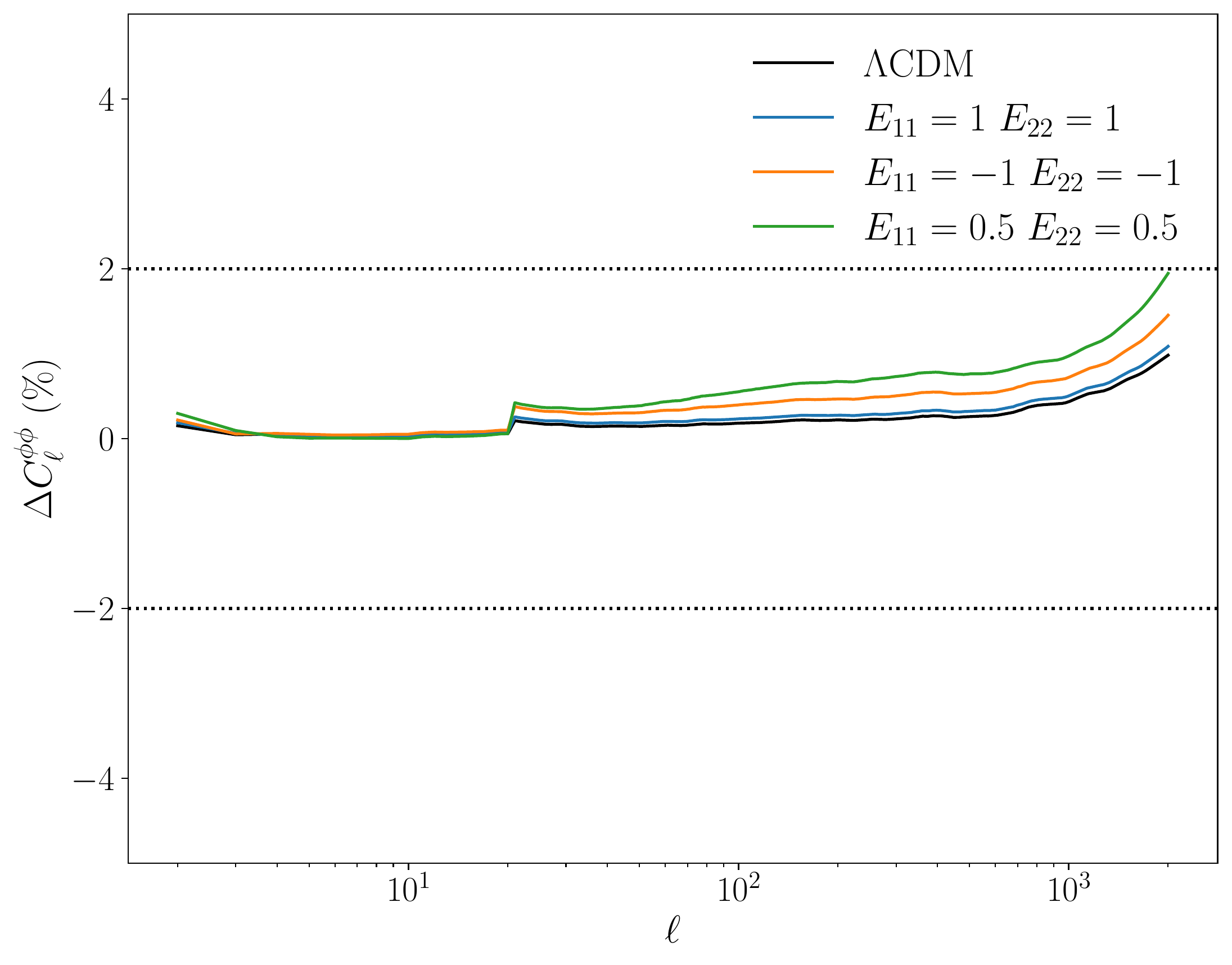}
	\end{tabular}
	\caption{Relative difference between the CMB temperature and lensing angular power spectrum produced by \texttt{MGCLASS II}, with respect to \texttt{MGCAMB}, considering the Planck late parameterization for different values of $E_{11}$ and $E_{22}$, while fixing the standard cosmological parameters to the Planck 2018 means.}
	\label{fig:comparison_spectra} \end{figure}

In order to validate our implementation of the modified evolution equations for background and perturbations, we compared the predictions of \texttt{MGCLASS II} with those obtained from \texttt{MGCAMB}. In \autoref{fig:comparison_spectra} we show the relative difference in the CMB temperature (left panel) and lensing potential (right panel) power spectra obtained using the Planck late parameterization described in \autoref{sec:parameterizations}. This figure shows how the two codes agree for nearly all scales below $1\%$, with the discrepancy reaching at most $2\%$ at very low multipoles for the temperature correlation, while the lensing potential power spectrum only differs of more than $1\%$ only at very high multipoles, even though always below $2\%$. In order to achieve this agreement, we used for the two Boltzmann solvers the precision settings described in Appendix C of \cite{Bellini:2017avd}.

\begin{figure}[!h]
	\centering
	\includegraphics[width=0.85\columnwidth]{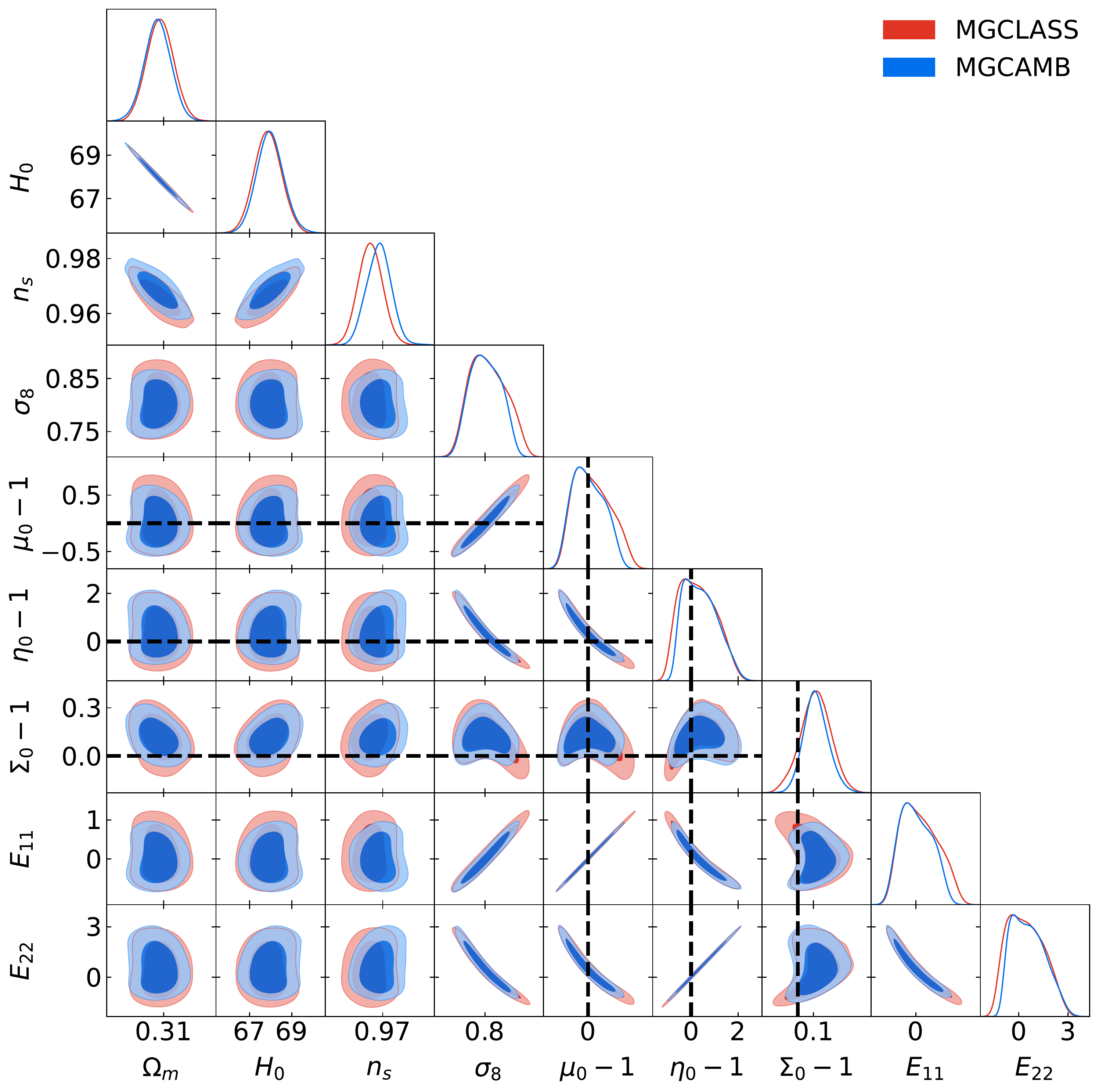}
	\caption{$68 \%$ and $95 \%$ confidence level contours obtained fitting the Planck 2018 data for temperature, polarization and CMB lensing using theoretical predictions from \texttt{MGCLASS II} (red contours) and \texttt{MGCAMB} (blue contours).}\label{fig:comparison_params}
\end{figure}

We show instead in \autoref{fig:comparison_params} the results obtained fitting the Planck 2018 data with \texttt{MGCLASS II} (red contours) and \texttt{MGCAMB} (blue contours), finding a good agreement on most cosmological parameters\footnote{discrepancies in the tails of the distribution for the $E_{11}$ and $E_{22}$ are probably due to the different level of convergence of the two runs.}. While this comparison is only done for Planck late, this choice only defines the specific trends of the MG functions, but uses the modified evolution equations of all the models implemented in \texttt{MGCLASS II}, thus ensuring that these are well implemented. Indeed, bugs can still be present in the implementation of the specific MG functions for the other models. While we compared the spectra computed by \texttt{MGCLASS II} and \texttt{MGCAMB} for the other parameterizations and models in common, obtaining similar results to those shown in \autoref{fig:comparison_params}, a more systematic comparison with external codes, also at the level of observational constraints, is left for a future publication.
 
\section{Constraints from cosmological data}\label{sec:analysis}

We want now to constrain the parameterizations discussed in \autoref{sec:parameterizations} using cosmological data; we focus here on CMB temperature, polarization and lensing potential spectra obtained from Planck 2018 \cite{Planck:2018nkj}, as done in the MG analysis of \cite{Aghanim:2018eyx}. We use the sampling and statistical modelling codes \texttt{Cobaya} \cite{Torrado:2020dgo} and \texttt{MontePython} \cite{2013JCAP...02..001A} to explore through a Monte Carlo Markov Chain (MCMC) Metropolis-Hastings (MH) sampler \cite{Lewis:2002ah,Lewis:2013hha} the space of free parameters, namely the baryonic and dark matter physical densities $\Omega_b h^2$ and $\Omega_ch^2$, the Hubble parameter $H_0$, the amplitude $A_s$ and tilt $n_s$ of the primordial perturbation power spectrum, the optical depth at reionization $\tau$ and, when moving away from the $\Lambda$CDM model, the parameters describing the MG functions we introduced in \autoref{sec:parameterizations}. At each point of the parameter space, \texttt{Cobaya} and \texttt{MontePython} obtain the theoretical predictions from \texttt{MGCLASS II} and compare them with the Planck data. For each point of the MCMC chains, we obtain as derived parameters the value of the three MG functions $\mu$, $\eta$ and $\Sigma$ in a set of equispaced redshift, in order to reconstruct their trends in time.

\subsection{$\Lambda$CDM background}\label{sec:LCDM_bkg}

As a first step, we neglect the contributions that the MG functions bring to the evolution of the background expansion; we assume that the latter is well described by a $\Lambda$CDM expansion history, and we focus on the impact that deviations from GR in the evolution of perturbations have on observables. 

\begin{figure}[!t]
	\centering
	\includegraphics[width=0.75\columnwidth]{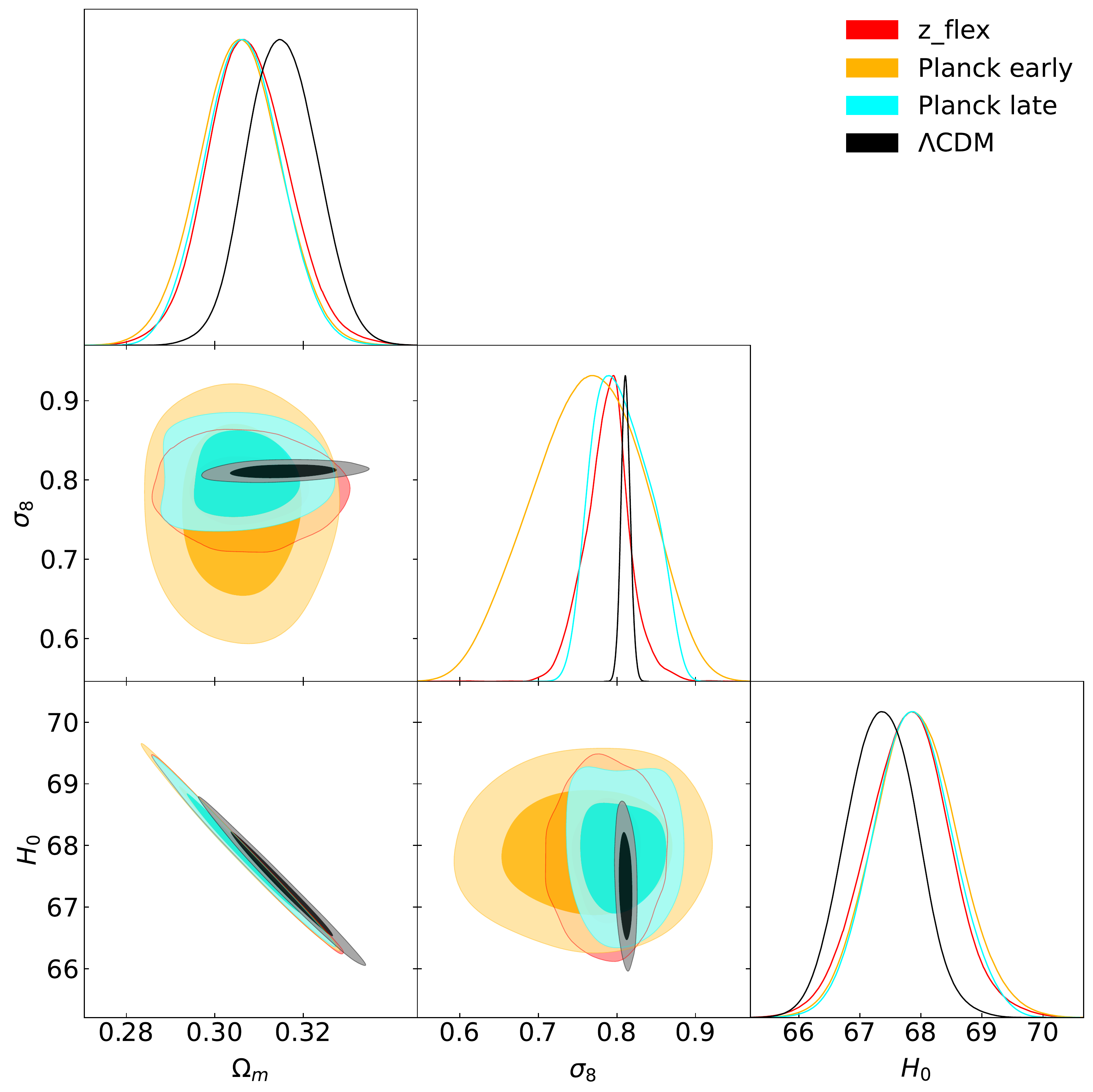}
	\begin{tabular}{cc}
	\includegraphics[width=0.45\columnwidth]{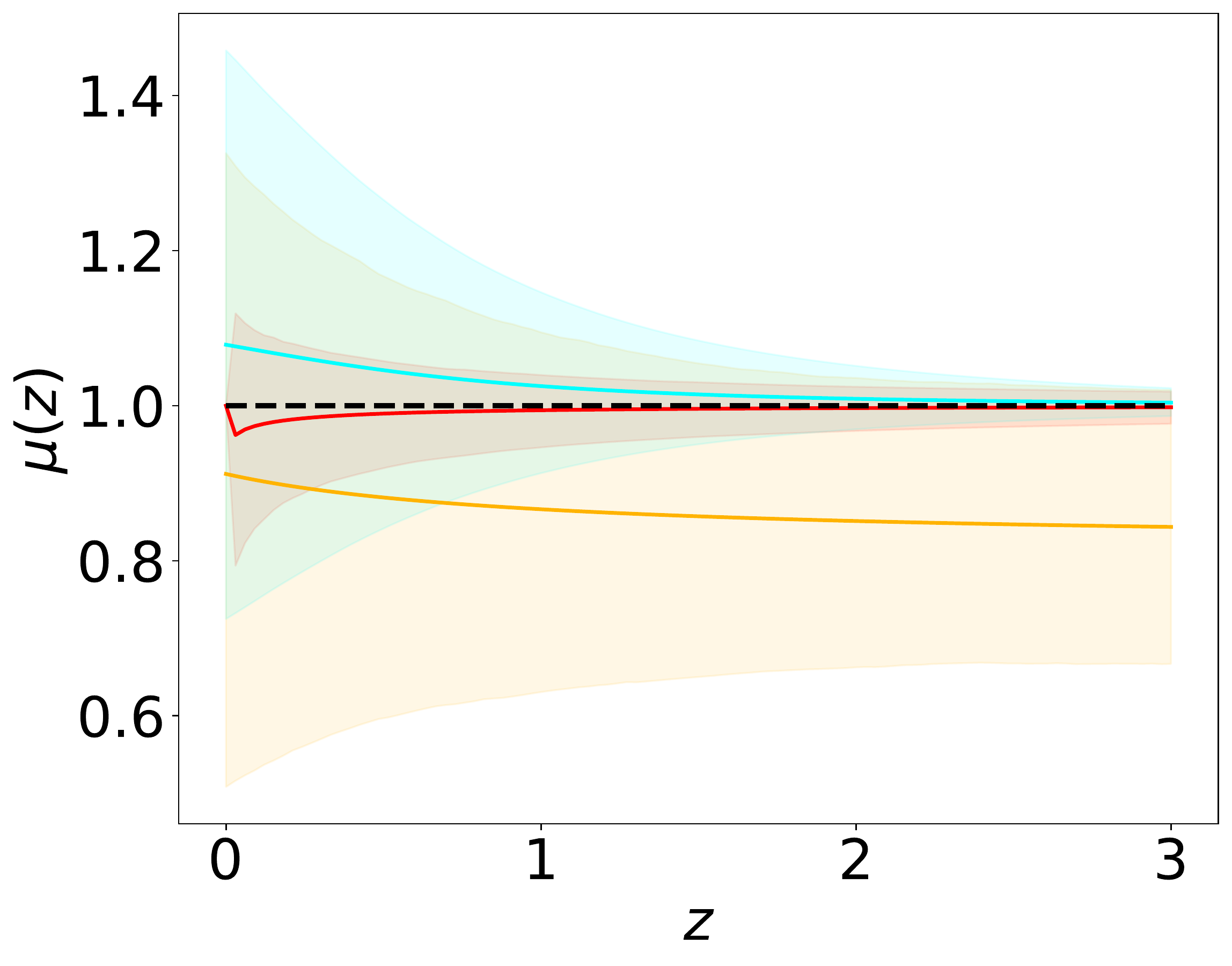} &
	\includegraphics[width=0.45\columnwidth]{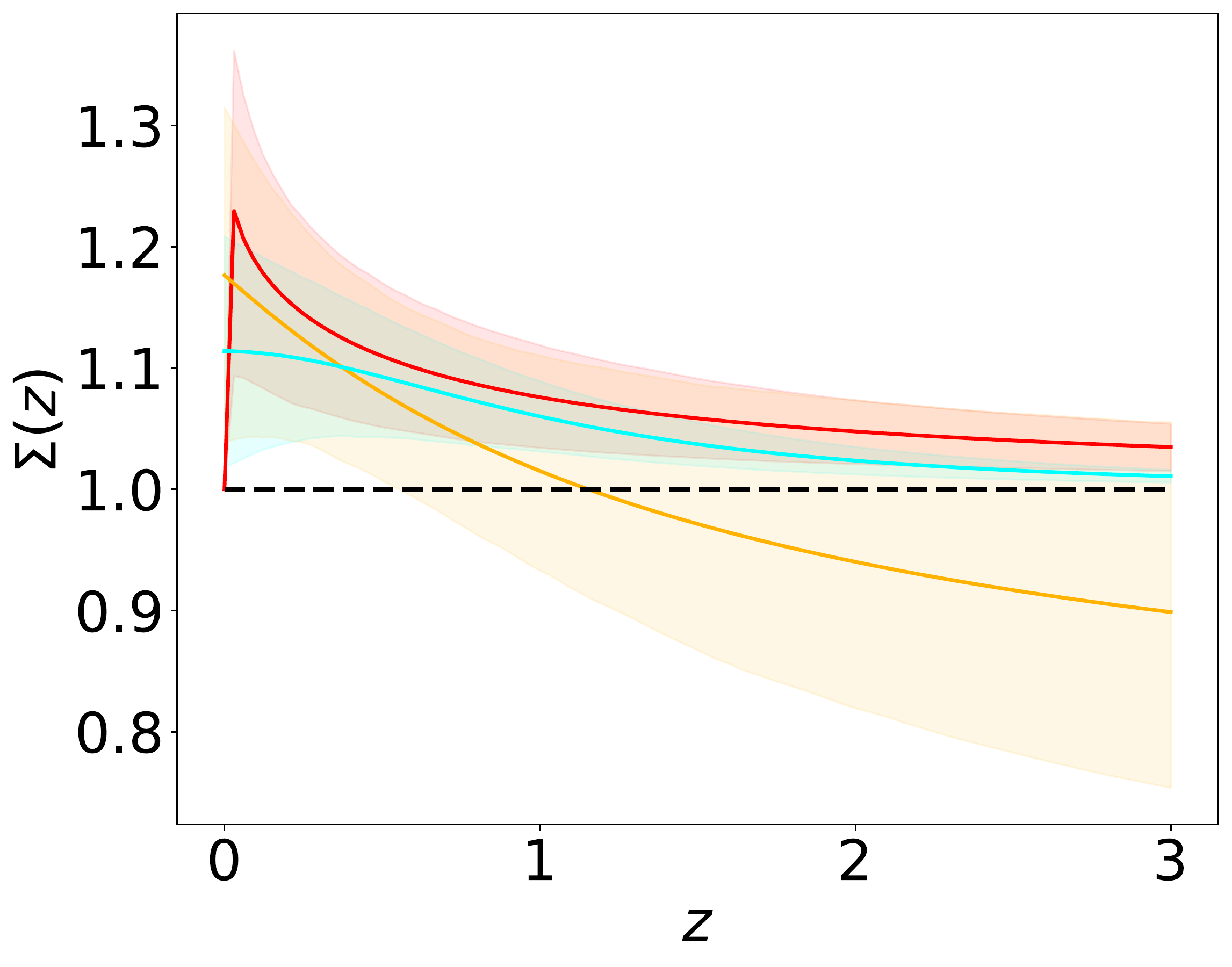}
	\end{tabular}
	\caption{Comparison between the constraints obtained assuming $\Lambda$CDM (black) and the MG parameterizations {\it Planck late} (cyan), {\it Planck early} (yellow) and {\it z\_flex late} (red). The bottom panels show the different reconstructions of $\mu(z)$ (left) and $\Sigma(z)$ (right), while the top panel show the results obtained on the standard cosmological parameters.}\label{fig:LCDMbkg}
\end{figure}

In \autoref{fig:LCDMbkg} we compare the results obtained using the two Planck-like parameterizations (late in purple and early in yellow) with those achieved assuming z\_flex with all three parameters ($g_\mu$, $g_\eta$, $n$) set free. The bounds obtained on the cosmological and MG parameters are shown in \autoref{tab:LCDMbkg}. We show both the two-dimensional contours on a subset of the standard parameters ($\Omega_{\rm m}$, $H_0$ and $\sigma_8$) and the reconstruction of the MG functions $\mu(z)$ and $\Sigma(z)$ obtained from the constraints on the parameters for each parameterization. One can notice how the behaviour of the Planck late and z\_flex parameterizations is extremely similar; both of them return to GR values at high redshifts, the first because of the decrease of $\Omega_{\rm DE}(z)$ and the second by construction, while their behaviour is different at late times, although the functions reconstructed at low redshifts are still compatible with each other. Moreover, in both cases we find that $\mu(z)$ is compatible and very close to the GR limit $\mu=1$, while larger deviations are allowed for $\Sigma$, with its GR limit outside the $68\%$ confidence level region for this function. This is not unexpected since, as pointed out in \cite{Aghanim:2018eyx}, it is the lensing effect on the CMB spectra that slightly favours modifications of gravity.

Given the similar behaviour of the MG functions, it is no surprise to see that also the constraints on the standard cosmological parameters behave similarly; in both cases the mean values for $\Omega_{\rm m}$ and $\sigma_8$ are unchanged with respect to $\Lambda$CDM (black contours), with their errors getting larger because of the additional free parameters, while $H_0$ central value is slightly shifted towards higher value, an effect that slightly reduces the tension between the CMB estimate and the local measurements of this parameter (see e.g. \cite{Perivolaropoulos:2021jda}). The Planck early parameterization behaves instead quite differently, due to the specific shape of the $\mu$ and $\eta$ functions assumed in \autoref{eq:plk_early}. On top of this, using this parameterization also leads to larger errors on both MG and cosmological parameters, given the higher number of free parameters with respect to the other two parameterizations considered. Overall, however, the hint for a $\Sigma(z)>1$ at small redshift is preserved also in this parameterization, with $\mu(z)$ that is compatible with the GR limit.

\begin{table}[]
    \centering
\begin{tabular} { l | c c c | c}
\noalign{\vskip 3pt}\hline\noalign{\vskip 1.5pt}\hline
                     &  Planck early             &  z\_flex                        &  Planck late                 &  $\Lambda$CDM\\
\cline{2-5}

 Parameter           &  68\% limits              &  68\% limits                    &  68\% limits                 &  68\% limits\\
\hline
$\Omega_{\rm c} h^2$ & $0.1185\pm 0.0015$        & $0.1188\pm 0.0017$              & $0.1186\pm 0.0014$           & $0.1200\pm 0.0012$\\
$\Omega_{\rm b} h^2$ & $0.02241\pm 0.00017$      & $0.02241^{+0.00016}_{-0.00018}$ & $0.02241\pm 0.00015$         & $0.02237\pm 0.00014$\\
$\ln10^{10}A_{\rm s}$& $3.027\pm 0.017$          & $3.032\pm 0.018$                & $3.027^{+0.017}_{-0.014}$    & $3.045\pm 0.014$\\
$n_{\rm s}$          & $0.9664\pm 0.0047$        & $0.9646^{+0.0061}_{-0.0055}$    & $0.9654\pm 0.0045$           & $0.9648\pm 0.0039$\\
$\tau_{\rm reio}$    & $0.0488\pm 0.0083$        & $0.0505^{+0.0086}_{-0.0072}$    & $0.0488^{+0.0083}_{-0.0067}$ & $0.0544\pm 0.0073$\\
$\sigma_8$           & $0.764^{+0.073}_{-0.065}$ & $0.786\pm 0.040$                & $0.806^{+0.030}_{-0.042}$    & $0.8111\pm 0.0058$\\
$H_0$                & $67.91\pm 0.68$           & $67.82\pm 0.78$                 & $67.87\pm 0.63$              & $67.37\pm 0.53 $\\
$\Omega_{\rm m}$     & $0.3057\pm 0.0090$        & $0.307\pm 0.010$                & $0.3064\pm 0.0084$           & $0.3151\pm 0.0072$\\
\hline
$E_{11 }$            & $-0.08^{+0.33}_{-0.45}$   & $---$                           & $0.11^{+0.38}_{-0.61}$       & $---$\\
$E_{22 }$            & $-1.34^{+2.2}_{-0.82}$    & $---$                           & $0.42^{+0.90}_{-1.4}$        & $---$\\
$E_{21 }$            & $1.16^{+0.70}_{-1.9}$     & $---$                           & $---$                        & $---$\\
$E_{12 }$            & $-0.09^{+0.49}_{-0.20}$   & $---$                           & $---$                        & $---$\\
$g_\mu$              & $---$                     & $-0.68^{+2.3}_{-0.97}$          & $---$                        & $---$\\
$g_\eta$             & $---$                     & $7.2^{+2.0}_{-9.4}$             & $---$                        & $---$\\
$n$                  & $---$                     & $0.125^{+0.013}_{-0.12}$        & $---$                   & $---$\\
\hline
\end{tabular}
\caption{Mean values and $68\%$ confidence level limits for cosmological and MG parameters obtained fitting the Planck late, Planck early and z\_flex parameterizations to the CMB data of Planck.}
\label{tab:LCDMbkg}
\end{table}

\subsubsection{Fixing $n$}

The parameterization we propose in this paper includes a parameter ($n$) ruling the quickness with which the MG function return to the baseline value after being allowed to vary. In the previous analysis we have kept this parameter free to vary, to investigate the full parameterization. However, having this parameter as free might create some issues: on the one hand the MG functions have two separate $\Lambda$CDM limits, vanishing $g_\mu$ or $g_\eta$ and $n=0$, and this creates posterior distributions that are complicated to sample, on the other hand the limit $n=0$ lies at the border of our prior choice and is therefore not perfectly explored by the MCMC. 

For such reasons, we explore here the case in which this parameter is kept fixed to unity, as an example case for which this parameterization shows deviations from GR at intermediate redshifts in the $\mu$ and $\eta$ functions. In the rest of the paper we will limit ourselves to this $n=1$ case to explore this parameterization in a limit where the posterior distributions are well behaved.

\begin{figure}[!t]
	\centering
	\includegraphics[width=0.75\columnwidth]{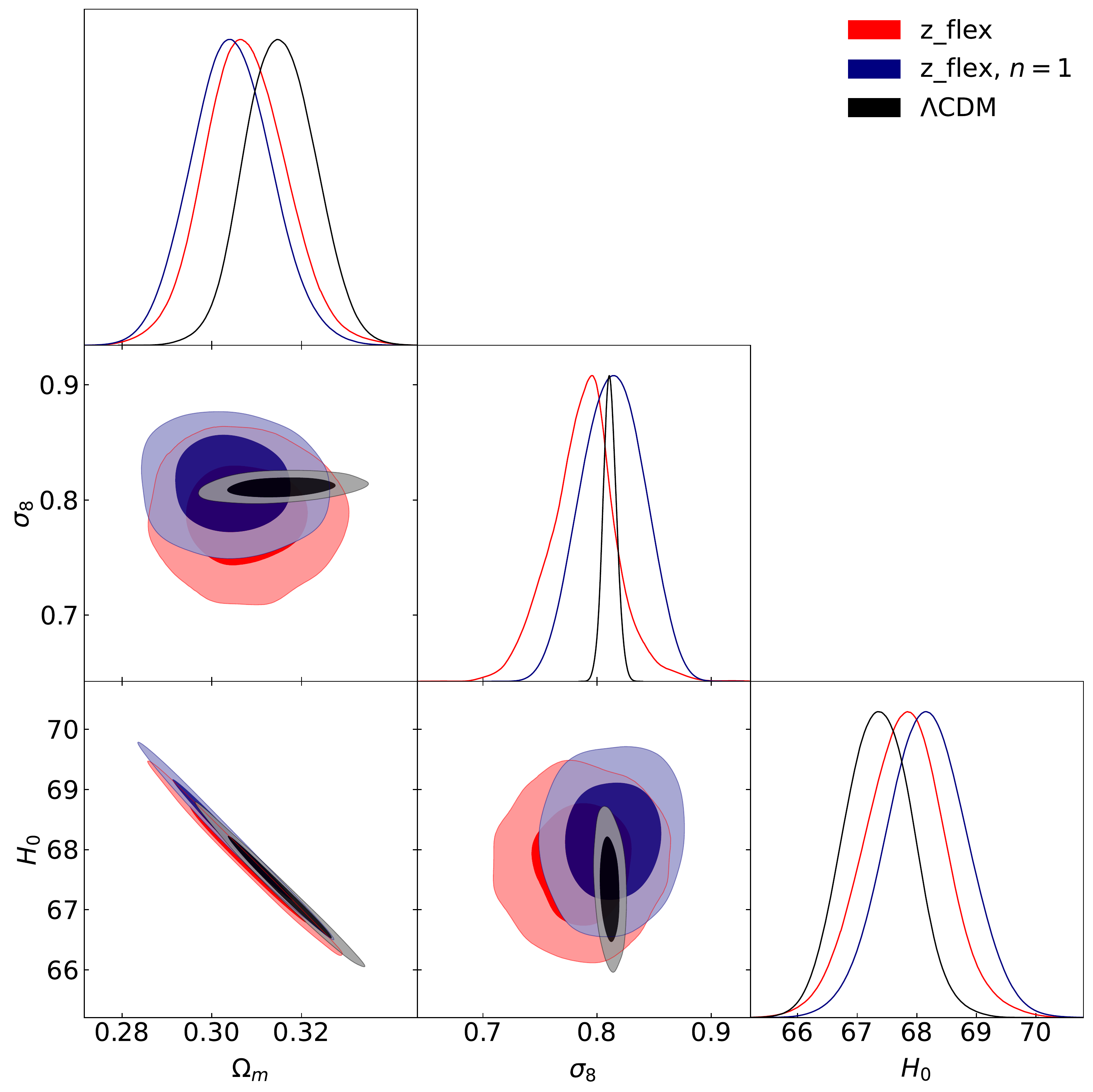}
	\begin{tabular}{cc}
	\includegraphics[width=0.45\columnwidth]{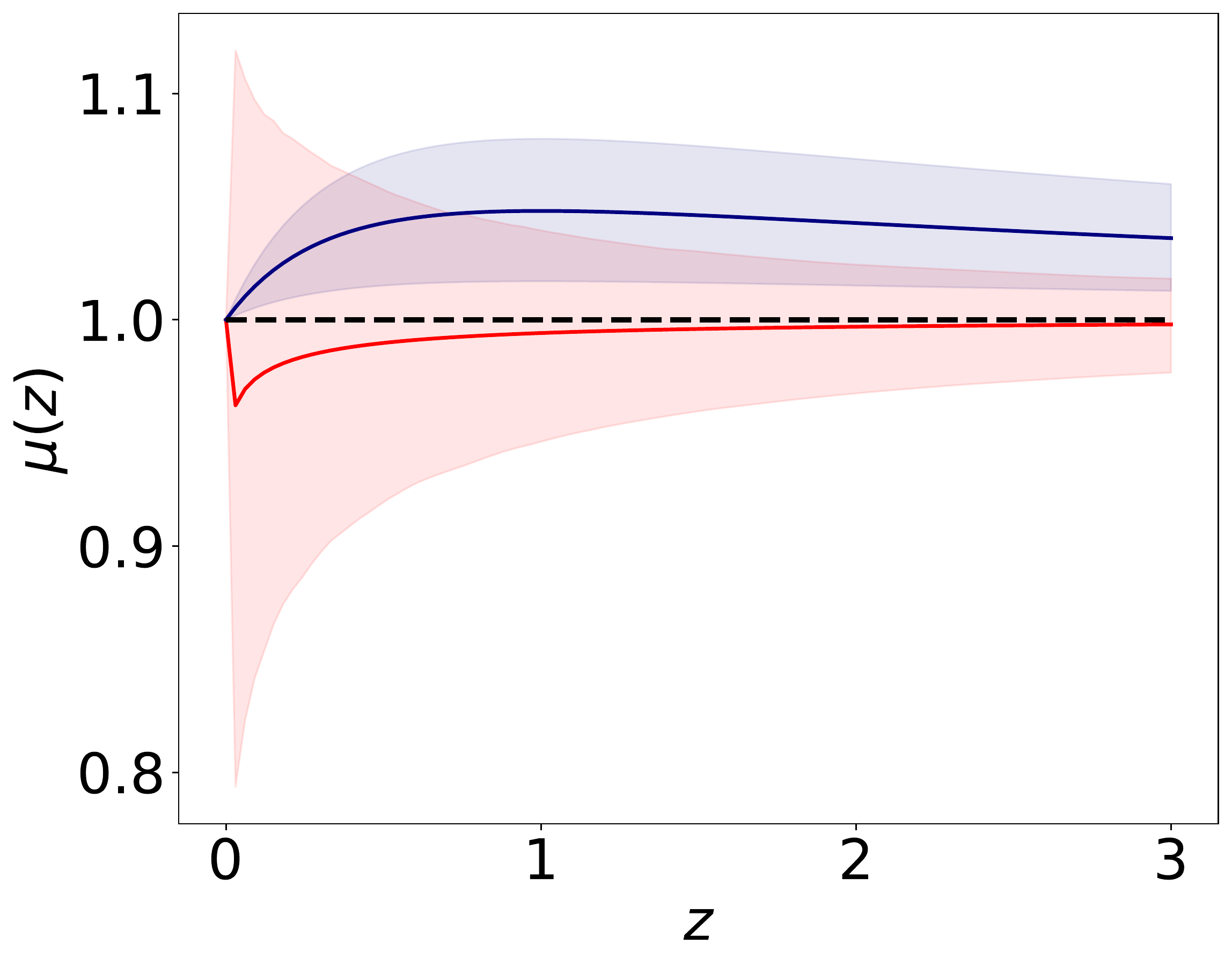} &
	\includegraphics[width=0.45\columnwidth]{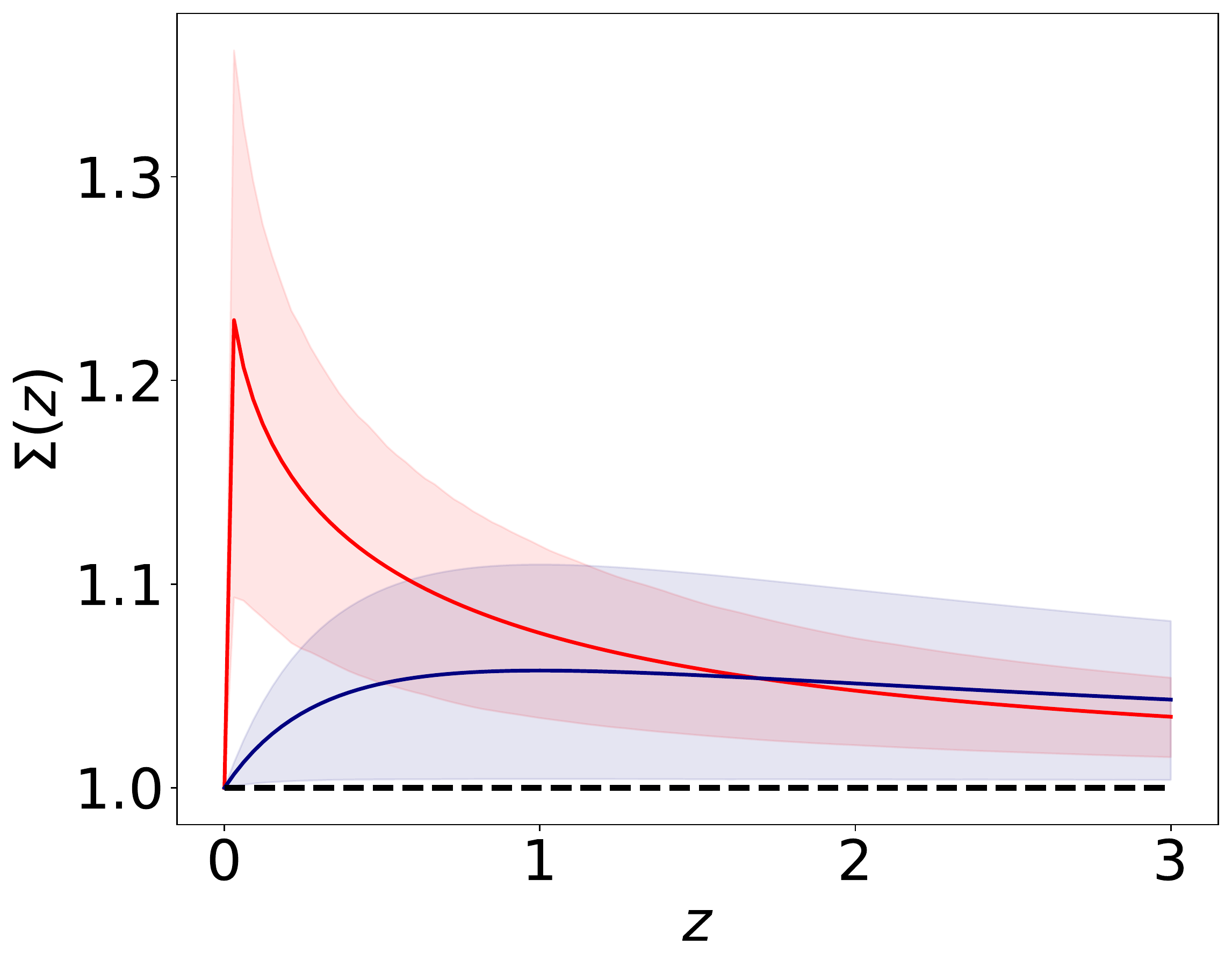}
	\end{tabular}
	\caption{Comparison between the constraints obtained assuming $\Lambda$CDM (black) and the MG parameterizations z\_flex, in the case of a free $n$ (red) and fixing $n=1$ (blue). The bottom panels show the different reconstructions of $\mu(z)$ (left) and $\Sigma(z)$ (right), while the top panel show the results obtained on the standard cosmological parameters.}\label{fig:zxpans_nz_1}
\end{figure}

We show the results of this analysis in \autoref{tab:LCDMbkg_n1} and \autoref{fig:zxpans_nz_1} where it is possible to see how the constraints on cosmological parameters are not affected significantly by fixing $n=1$. This could be also inferred by looking at the bottom panels of \autoref{fig:zxpans_nz_1} where, despite the different mean behaviour of the two functions for free and fixed $n$, $\mu$ and $\Sigma$ appear in agreement at less than $1\sigma$, thus highlighting how CMB alone is not able to significantly distinguish between such cases. At the level of $1\sigma$, however, it is possible to see how both $\mu(z)$ and $\Sigma(z)$ deviate from the GR limit for $n=1$, while for a free $n$, $\mu(z)$ was consistent with unity. Indeed, for this specific choice of $n$ we find that $\mu(z)\approx\Sigma(z)$, which is a condition that arises for stable Horndeski theories of gravity \cite{Espejo:2018hxa}.

\begin{table}[]
    \centering
\begin{tabular} { l | c c | c}
\noalign{\vskip 3pt}\hline\noalign{\vskip 1.5pt}\hline
                      &  z\_flex                        &  z\_flex ($n=1$)       &  $\Lambda$CDM\\
\cline{2-4}

 Parameter            &  68\% limits                    &  68\% limits           &  68\% limits\\
\hline
$\Omega_{\rm c} h^2$  & $0.1188\pm 0.0017$              & $0.1182\pm 0.0015$     & $0.1200\pm 0.0012          $\\
$\Omega_{\rm b} h^2$  & $0.02241^{+0.00016}_{-0.00018}$ & $0.02250\pm 0.00016$   & $0.02237\pm 0.00014        $\\
$\ln10^{10}A_{\rm s}$ & $3.032\pm 0.018$                & $3.029^{+0.017}_{-0.015}$       & $3.045\pm 0.014            $\\
$n_{\rm s}$           & $0.9646^{+0.0061}_{-0.0055}$    & $0.9698\pm 0.0048$     & $0.9648\pm 0.0039          $\\
$\tau_{\rm reio}$     & $0.0505^{+0.0086}_{-0.0072}$    & $0.0495^{+0.0079}_{-0.0072}$     & $0.0544\pm 0.0073          $\\
$\sigma_8$            & $0.786\pm 0.040$                & $0.814\pm 0.027$       & $0.8111\pm 0.0058          $\\
$H_0$                 & $67.82\pm 0.78$                 & $68.16\pm 0.67$        & $67.37\pm 0.53             $\\
$\Omega_{\rm m}$      & $0.307\pm 0.010$                & $0.3045\pm 0.0089$     & $0.3151\pm 0.0072          $\\
\hline
$g_\mu$               & $-0.68^{+2.3}_{-0.97}$          & $0.19^{+0.12}_{-0.13}$         & $--$\\
$g_\eta$              & $7.2^{+2.0}_{-9.4}$             & $0.08^{+0.45}_{-0.57}$ & $--$\\
$n$                   & $0.125^{+0.013}_{-0.12}$        & $--$                   & $--$\\
\hline
\end{tabular}
\caption{Mean values and $68\%$ confidence level limits for cosmological and MG parameters obtained fitting the z\_flex parameterization to the CMB data of Planck, with a free $n$ parameter and fixing it to $n=1$.}
\label{tab:LCDMbkg_n1}
\end{table}

\subsubsection{Assuming no lensing deviation}

As pointed out in \cite{Aghanim:2018eyx}, the main responsible of deviations from the GR limit in the MG functions is the excess of lensing that can be found in Planck measured CMB power spectra. It is therefore of interest to understand if the hints for MG we found above are driven by such feature. The \texttt{MGCLASS} code allows to manually impose $\Sigma(k,z)=1$, reducing the free MG parameters to only those determining $\mu(k,z)$ \cite{Brando:2019xbv} and obtaining the third function as 

\begin{equation}
    \eta(k,z) = -1+\frac{2}{\mu(k,z)}\, .
\end{equation}

Notice that here we are not making any assumption on how such feature would be achievable in a theoretical model alternative to GR, but rather only impose this condition at the phenomenological level. There are however models for which this is expected, e.g. in $f(R)$ theories \cite{Hu:2007nk}.

\begin{figure}[!t]
	\centering
	\begin{tabular}{cc}
	\includegraphics[width=0.35\columnwidth]{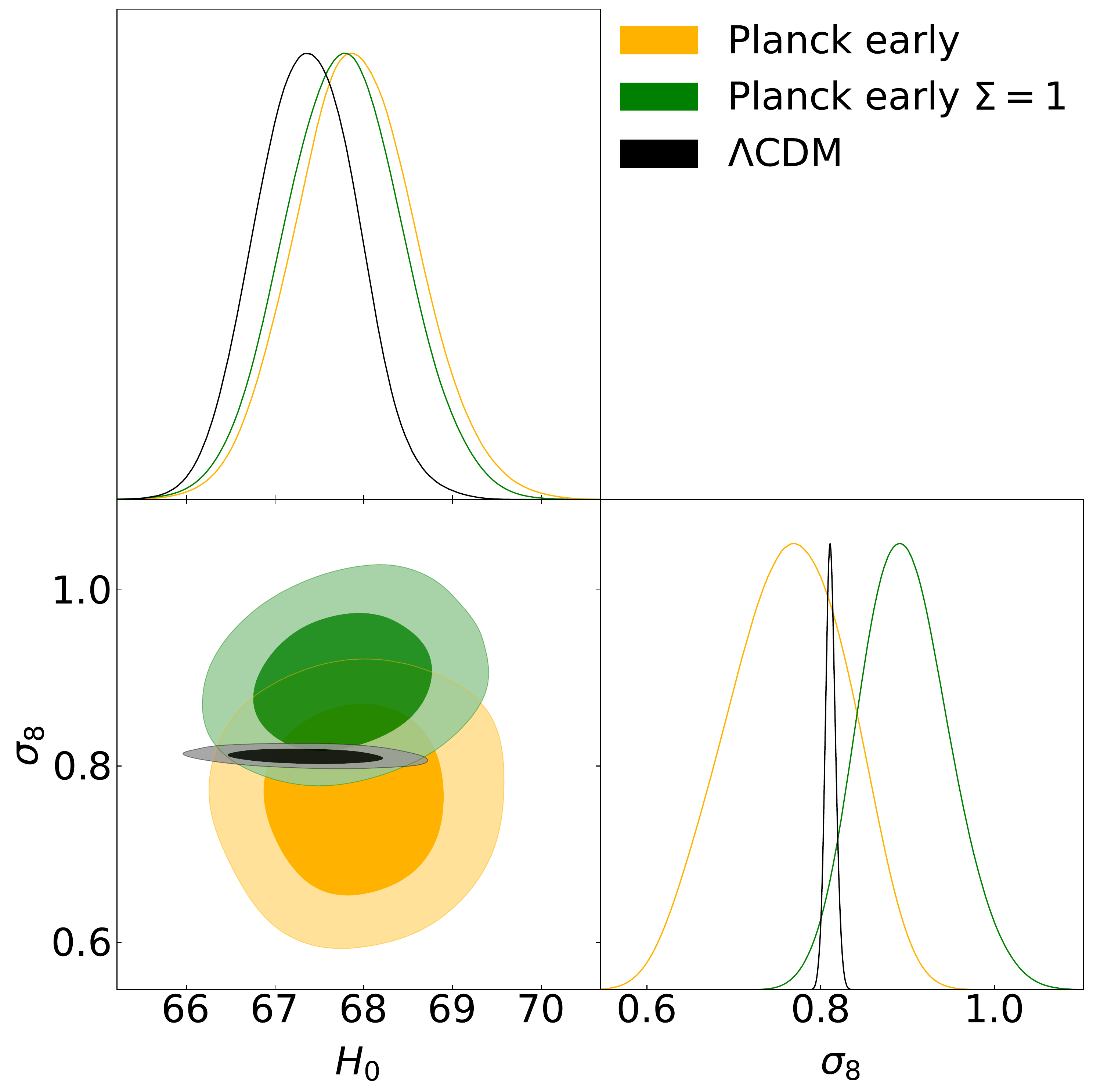} &
	\includegraphics[width=0.35\columnwidth]{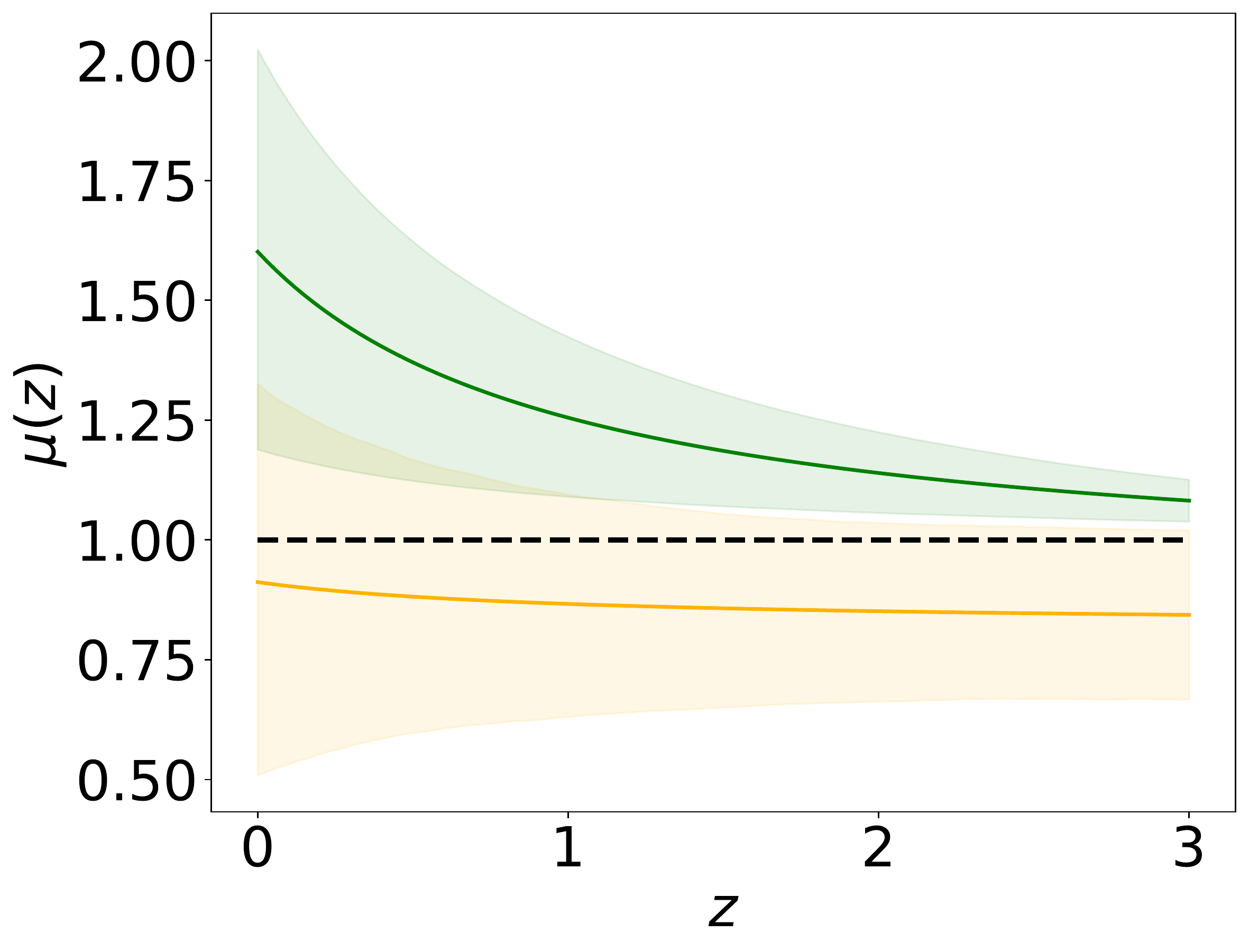} \\
	\includegraphics[width=0.35\columnwidth]{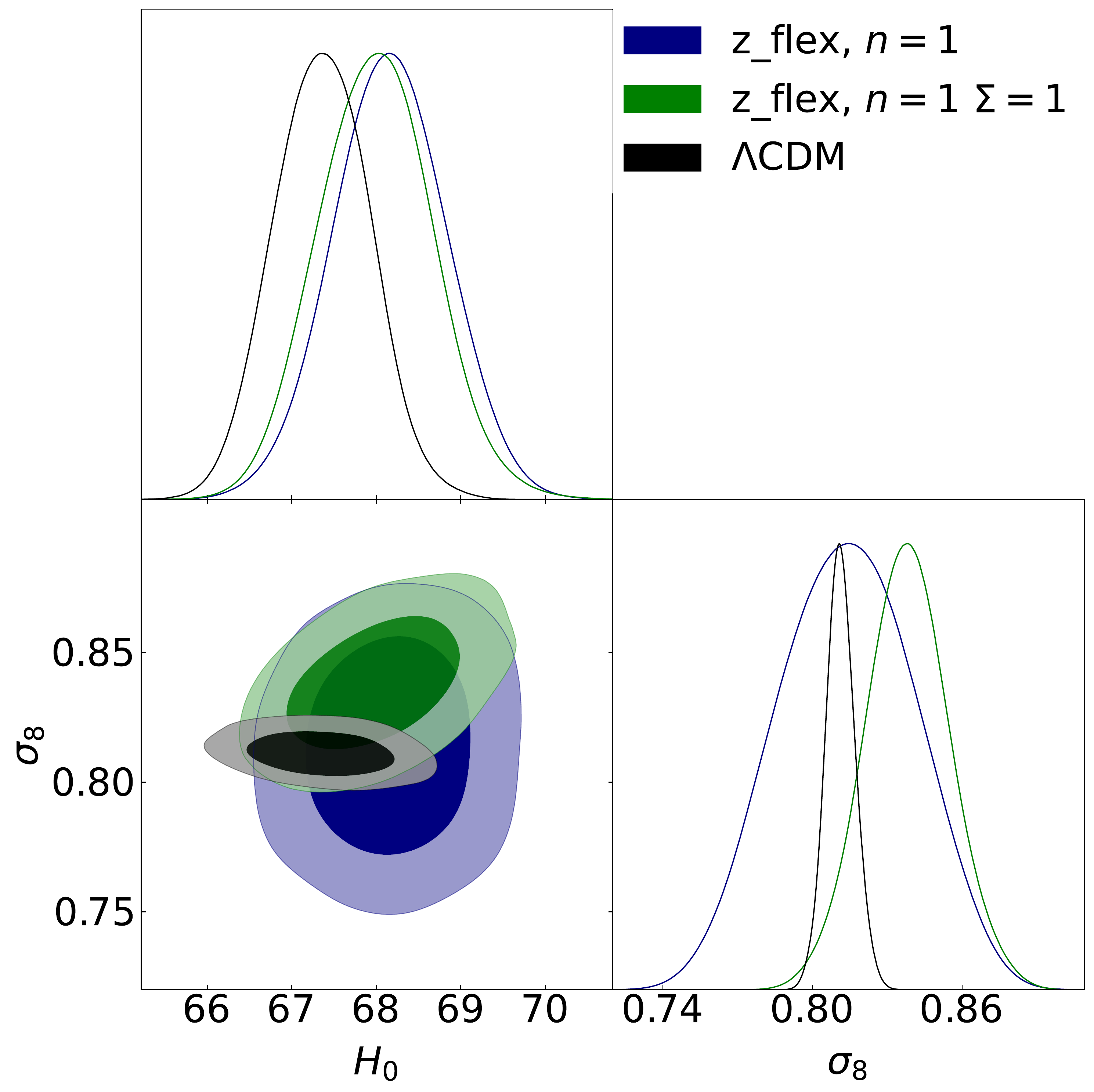} &
	\includegraphics[width=0.35\columnwidth]{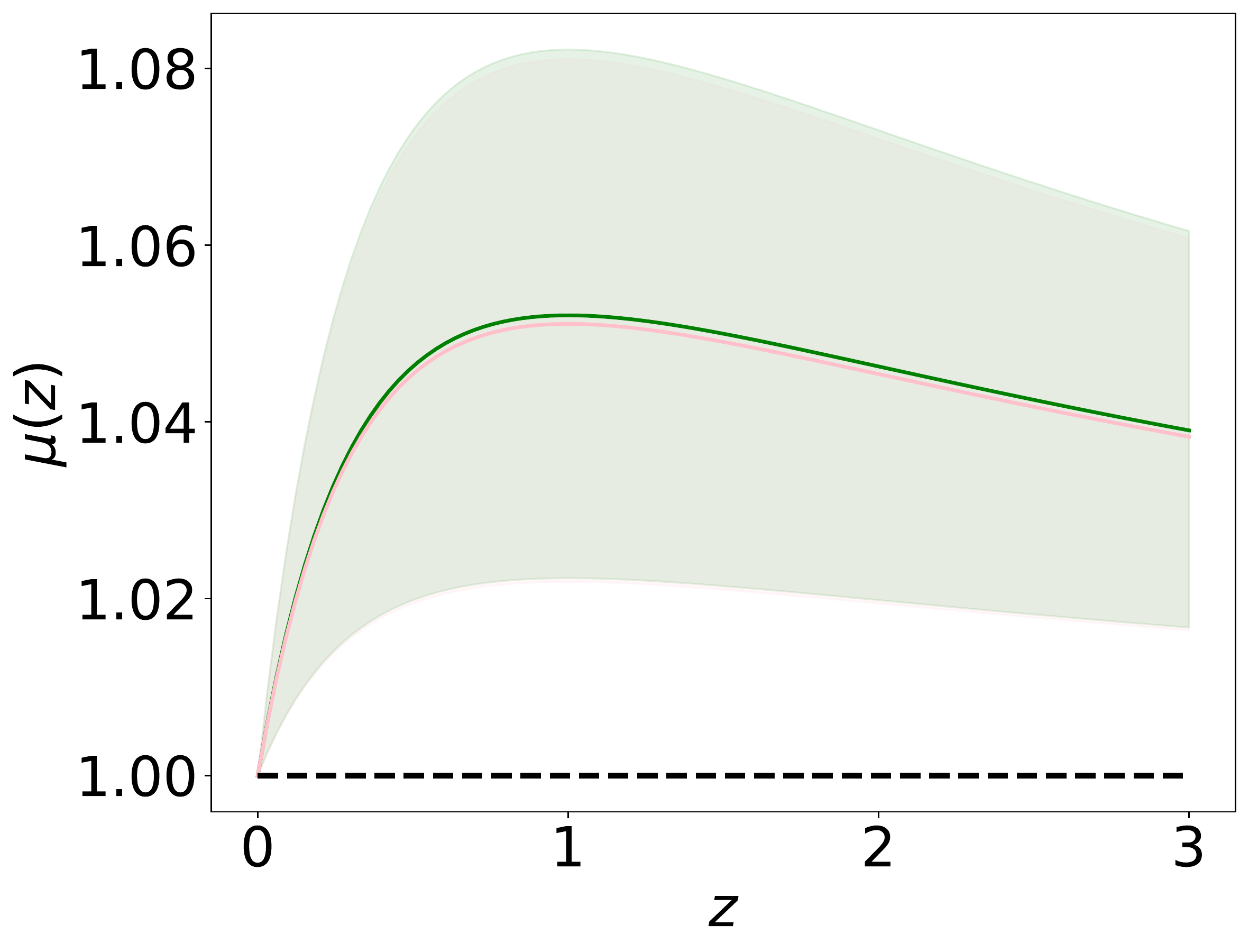} \\	
	\includegraphics[width=0.35\columnwidth]{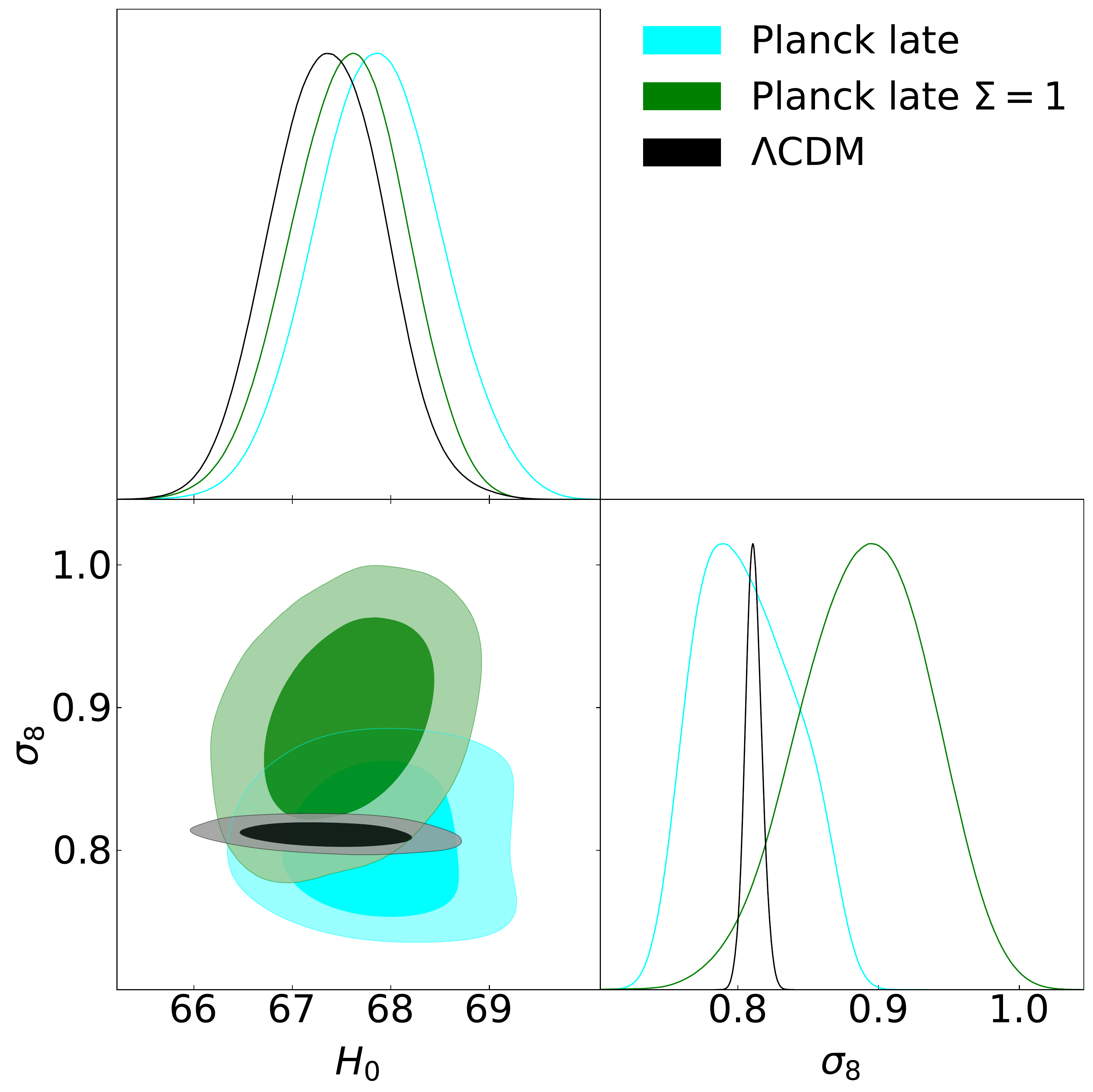} &
	\includegraphics[width=0.35\columnwidth]{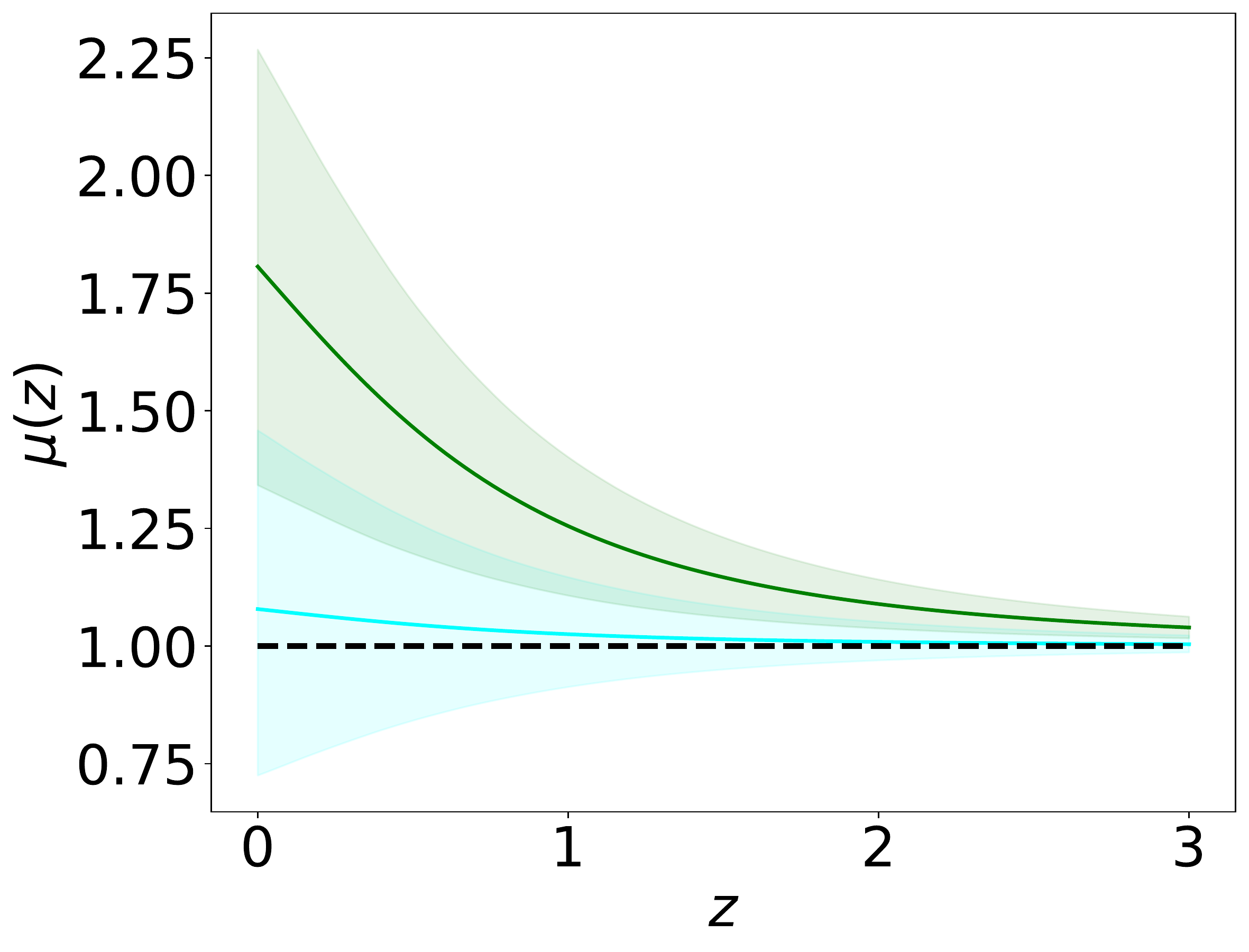} \\
	\end{tabular}
	\caption{Comparison between the constraints obtained on cosmological parameters and $\mu(z)$ when $\Sigma$ is assumed to be unity with the case of a free $\Sigma(z)$. The top panels show the comparison for Planck early, the central ones contain results for z\_flex ($n=1$), and the bottom panels refer to Planck late.}\label{fig:results_sig1}
\end{figure}

In \autoref{tab:LCDMbkg_sig1} and \autoref{fig:results_sig1} we show the impact of assuming $\Sigma=1$ on the constraints on cosmological parameters and $\mu(z)$. We find that for all three parameterizations, imposing $\Sigma=1$ requires an increase in the estimated values of $\mu(z)$. Such an effect is indeed related to the preference of CMB data for a lensing amplitude higher than the standard one; as the parameterizations are now not able to account for this with a direct modification of the lensing effect, the data drag $\mu$ to higher values, as this increases the amplitude of the matter power spectrum at low redshift, thus also increasing the lensing signal. In particular, we see how for both Planck early and Planck late, the mean of the $E_{11}$ parameter is shifted towards higher values, with the bound obtained in the latter parameterization that departs of $\approx2\sigma$ from the GR limit $E_{11}=0$. For what concerns z\_flex (with $n=1$), we see the same effect at play, although less significant, with an increased value of $g_\mu$ and the $\mu(z)$ function that is enhanced with respect to the case where $\Sigma$ is a free function.

The preference for an enhanced $\mu(z)$ with respect to the GR limit has however the consequence of increasing the recovered value of $\sigma_8$, as this is affected by the changes in the matter power spectrum, and therefore to worsen the tension between CMB and low redshift surveys on the measurement of this parameter (see e.g. \cite{Perivolaropoulos:2021jda}).

\begin{table}[]
    \centering
\begin{tabular} { l | c c c}
\noalign{\vskip 3pt}\hline\noalign{\vskip 1.5pt}\hline
                     &  Planck early ($\Sigma=1$)   &  z\_flex ($\Sigma=1$)       &  Planck late ($\Sigma=1$)\\
\cline{2-4}

 Parameter           &  68\% limits              &  68\% limits                 &  68\% limits\\
\hline
$\Omega_{\rm c} h^2$ & $0.1189\pm 0.0015$        & $0.1186\pm 0.0015$           & $0.1194\pm 0.0013$\\
$\Omega_{\rm b} h^2$ & $0.02240\pm 0.00016$      & $0.02247\pm 0.00016$         & $0.02237\pm 0.00015$\\
$\ln10^{10}A_{\rm s}$& $3.031^{+0.017}_{-0.015}$ & $3.033^{+0.017}_{-0.015}$    & $3.036\pm 0.015$\\
$n_{\rm s}$          & $0.9654\pm 0.0047$        & $0.9688\pm 0.0049$           & $0.9641\pm 0.0042$\\
$\tau_{\rm reio}$    & $0.0501\pm 0.0081$        & $0.0509^{+0.0082}_{-0.0073}$ & $0.0519\pm 0.0074$\\
$\sigma_8$           & $0.897^{+0.047}_{-0.055}$ & $0.838\pm 0.017$    & $0.891^{+0.049}_{-0.044}$\\
$H_0$                & $67.78\pm 0.66$           & $68.00\pm 0.67$              & $67.56\pm 0.58$\\
$\Omega_{\rm m}$     & $0.3076\pm 0.0089$        & $0.3066\pm 0.0089$           & $0.3106^{+0.0073}_{-0.0083}$\\
\hline
$E_{11 }$            & $0.60^{+0.38}_{-0.46}$    & $---$                        & $1.17^{+0.71}_{-0.59}$\\
$E_{22 }$            & $---$                     & $---$                        & $---$\\
$E_{21 }$            & $---$                     & $---$                        & $---$\\
$E_{12 }$            & $-0.69^{+0.55}_{-0.45}$   & $---$                        & $---$\\
$g_\mu$              & $---$                     & $0.21\pm 0.12$     & $---$\\
$g_\eta$             & $---$                     & $---$                        & $---$\\
\hline
\end{tabular}
\caption{Mean values and $68\%$ confidence level limits for cosmological and MG parameters obtained fitting the Planck late, Planck early and z\_flex ($n=1$) parameterizations to the CMB data of Planck, when no lensing deviation is assumed, i.e. $\Sigma=1$.}
\label{tab:LCDMbkg_sig1}
\end{table}

\subsection{MG background}\label{sec:MG_bkg}

While the assumption of a $\Lambda$CDM background is common in MG analyses (see e.g. \cite{Aghanim:2018eyx}) and it allows to focus the investigation on the impact of GR alternatives in the perturbation sector, we have shown in \autoref{sec:MGgeneralities} that a deviation of $\mu(z)$ and $\eta(z)$ from their GR value leads to modifications to the expansion history. Here we include such modifications in our analysis as described in \autoref{sec:MGgeneralities}; within the assumptions of QSA and of a minimally coupling scalar field, the MG functions also enter in the Friedmann equations and impact the background expansion of the Universe. We constrain here the Planck early and z\_flex parameterization assuming that they are describing a model for which such assumptions hold, and therefore explore how background information can constrain their parameters.\footnote{We do not analyze the Planck late parameterization in this context; the dependence of $\mu(z)$ and $\eta(z)$ on $\Omega_{\rm DE}(z)$ complicates the propagation of MG effects to the background unless we parameterize the model as function of a normalized $\Omega_{\rm DE}(z)/\Omega_{\rm DE}(z=0)$. Alhough we also implemented the latter in \texttt{MGCLASS II}, this does not corresponds to the 'late' model of Planck collaboration that we are using as one of our benchmarks.} For such reason, in this analysis we do not use only Planck CMB data, but also include background observations from BAO measurements \cite{Beutler:2011hx,Ross:2014qpa,BOSS:2016wmc}.

\subsubsection{Planck early parameterization}

\begin{figure}[!t]
	\centering
	\includegraphics[width=0.75\columnwidth]{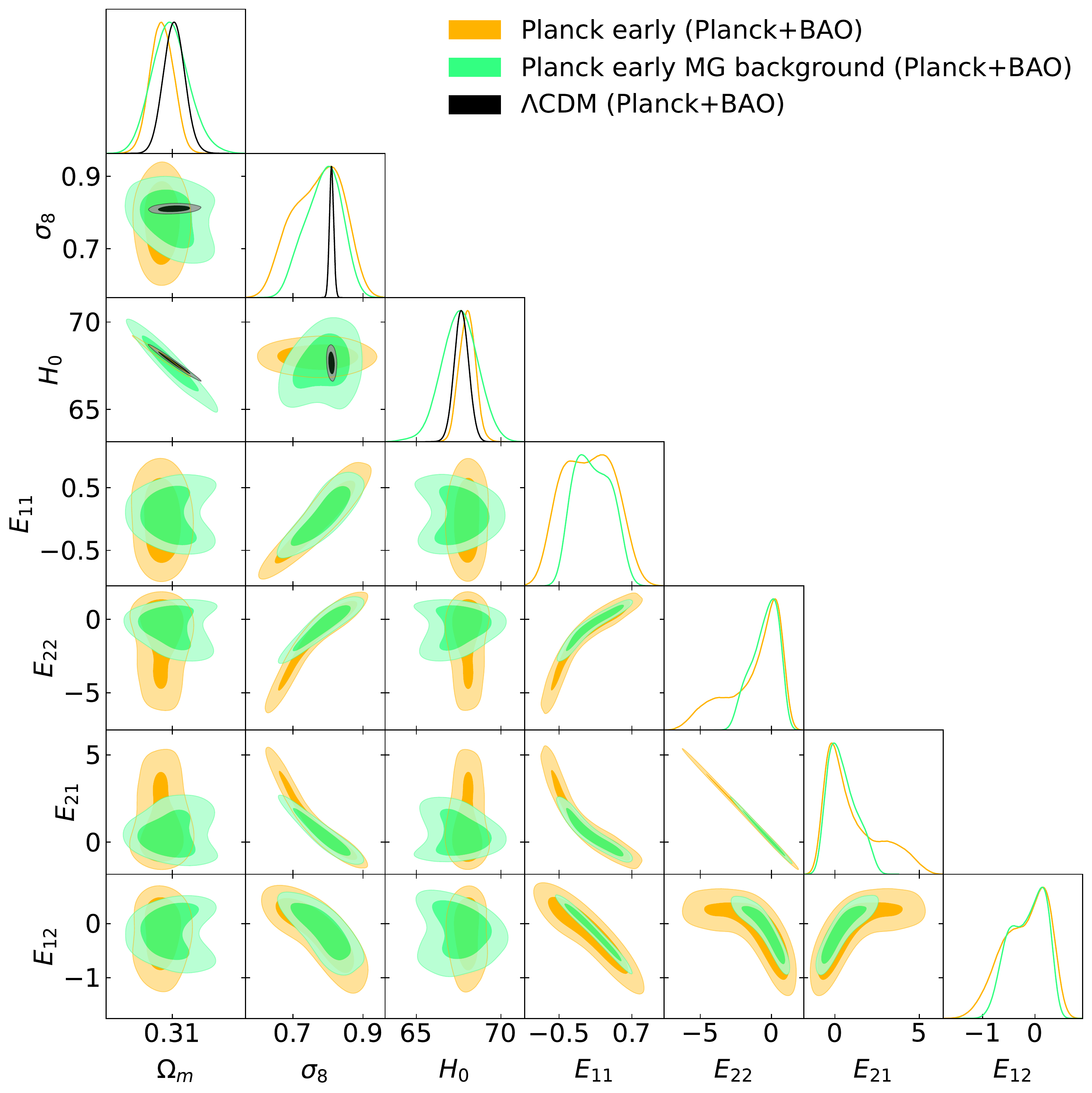}
	\begin{tabular}{cc}
	\includegraphics[width=0.45\columnwidth]{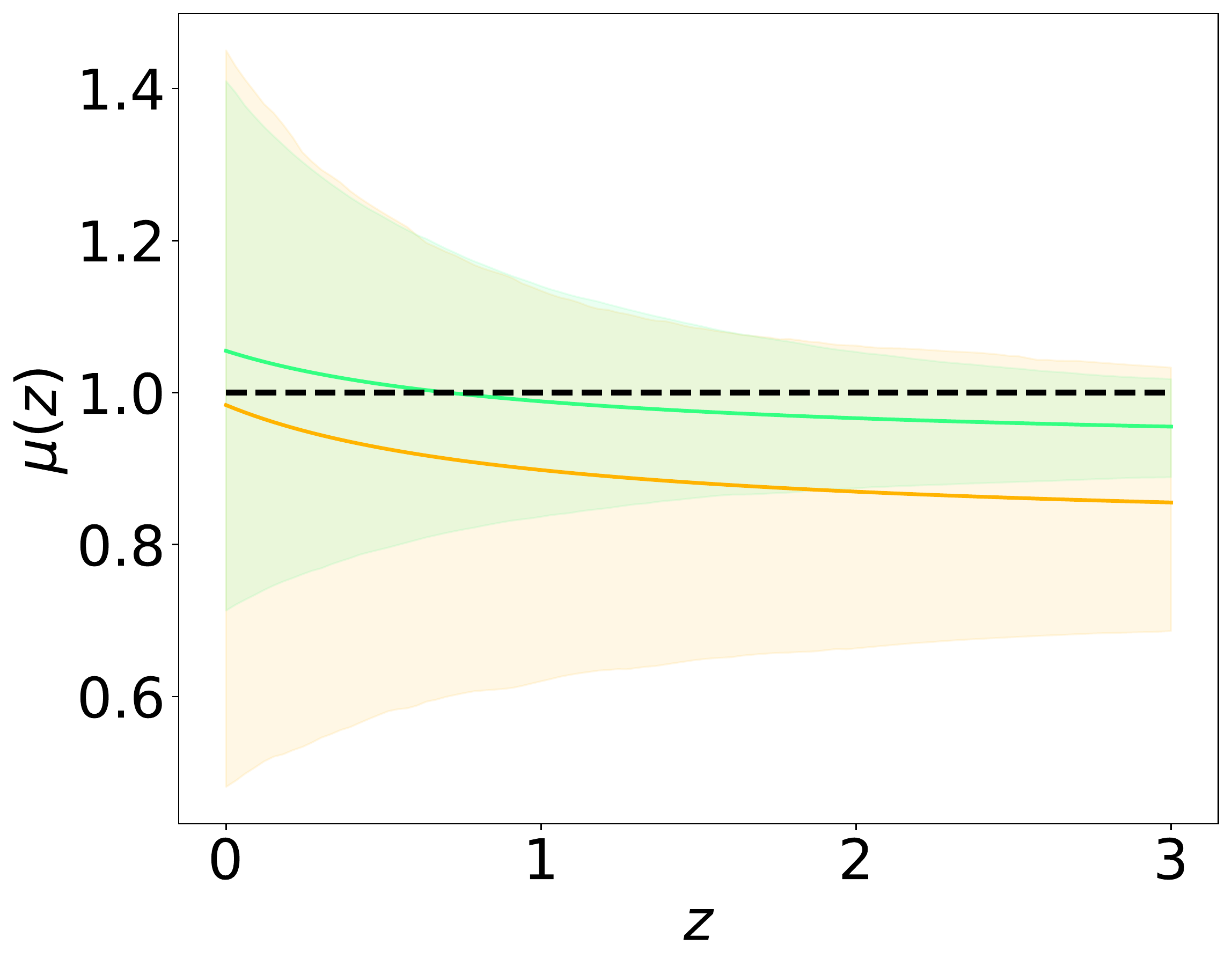} &
	\includegraphics[width=0.45\columnwidth]{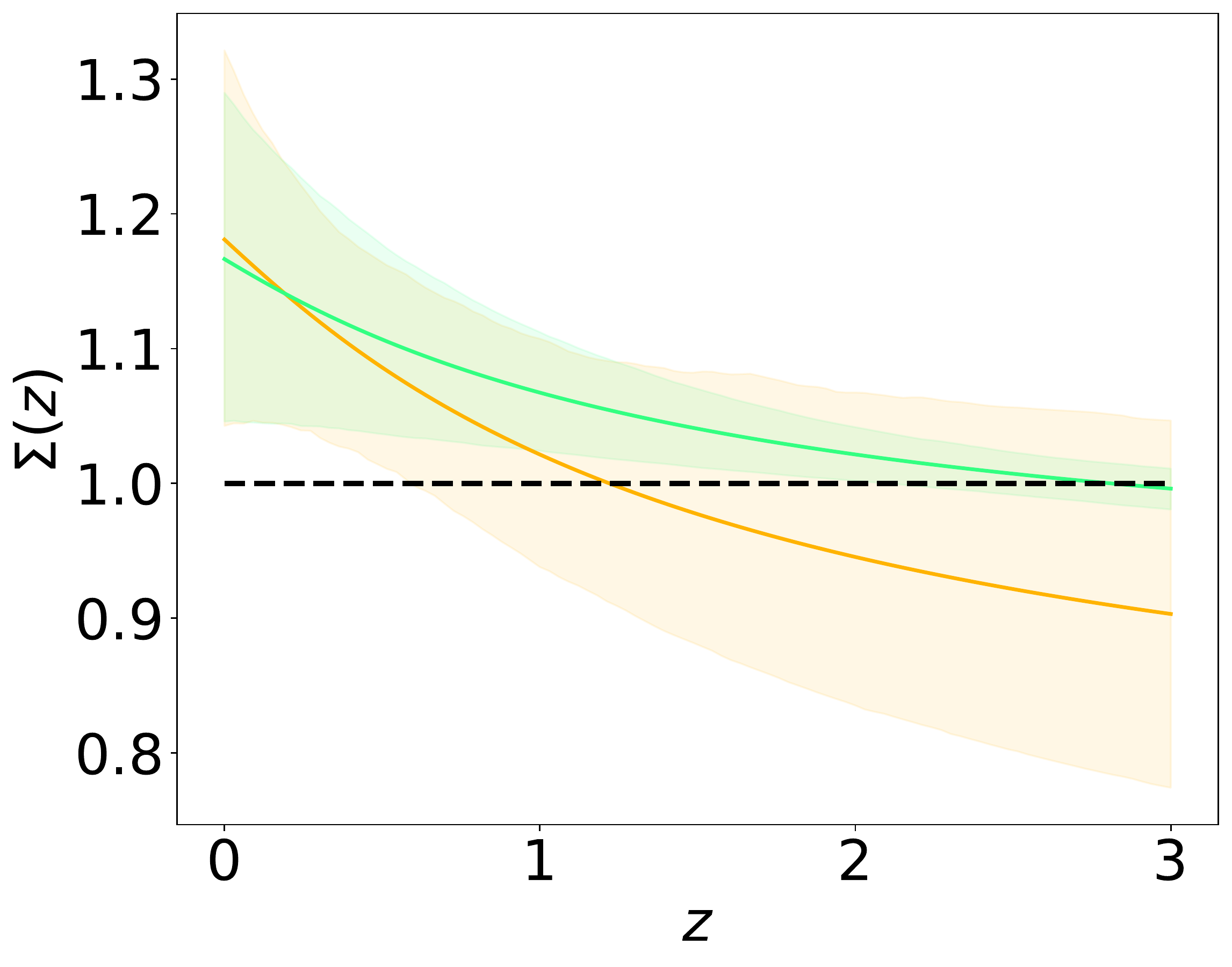}
	\end{tabular}
	\caption{Constraints on Planck early when its effect on the background expansion are considered (light green), compared with the case with a $\Lambda$CDM background (yellow) and with the $\Lambda$CDM limit (black). The bottom panel shows the constraints on the MG functions $\mu(z)$ and $\Sigma(z)$, while the top panel show the results obtained on the standard cosmological parameters. The data combination used here include CMB and BAO.}\label{fig:MGbkg_pkearly}
\end{figure}

We first focus on the simple Planck early parameterization, exploring how the inclusion of its effect on the background might affect the constraints. We show the results of this analysis in \autoref{tab:MGbkg_pkearly} and \autoref{fig:MGbkg_pkearly}; we can notice how the inclusion of the MG effects on the background leads to smaller deviations from GR in the $\mu(z)$ function, with its error reduced with respect to the case when the background is assumed to be $\Lambda$CDM. While the same reduction of the error is also present in $\Sigma(z)$, this function still departs from its GR limit, highlighting how the preference for a high lensing effect in CMB data is still significant in this setting.

If we look instead at the constraints on the standard cosmological parameters, we notice how including the MG effects on the background expansion leads to tighter constrain on $\sigma_8$, a perturbation parameter that now is also constrained by background observations through its dependence on the MG parameters, but we obtain at the same time looser constraints on other parameters, e.g. $H_0$; the latter effect is due to the fact that, while assuming $\Lambda$CDM we only have a weak degeneracy between MG parameters and $H_0$, this becomes stronger when MG parameters directly influence the expansion history.

\begin{table}[]
    \centering
\begin{tabular} { l | c c | c}
\noalign{\vskip 3pt}\hline\hline
 & \multicolumn{2}{c|}{Planck early} & $\Lambda$CDM\\
\hline
                     & $\Lambda$ background &  MG background    &  \\
\cline{2-4}

 Parameter           &  68\% limits              &  68\% limits                    &  68\% limits\\
\hline
$\Omega_{\rm c} h^2$ & $0.1183\pm 0.0010$        & $0.1186\pm 0.0015$              & $0.11933\pm 0.00093$\\
$\Omega_{\rm b} h^2$ & $0.02243\pm 0.00014$      & $0.02240^{+0.00015}_{-0.00016}$ & $0.02241\pm 0.00014$\\
$\ln10^{10}A_{\rm s}$& $3.026\pm 0.016$          & $3.029^{+0.014}_{-0.017}$       & $3.047\pm 0.015$\\
$n_{\rm s}$          & $0.9670\pm 0.0039$        & $0.9662\pm 0.0049$              & $0.9663\pm 0.0037$\\
$\tau_{\rm reio}$    & $0.0492\pm 0.0077$        & $0.0496^{+0.0069}_{-0.0082}$    & $0.0561\pm 0.0074$\\
$\sigma_8$           & $0.767^{+0.087}_{-0.068}$ & $0.787^{+0.059}_{-0.046}$       & $0.8103\pm 0.0060$\\
$H_0$                & $68.05\pm 0.47$           & $67.6\pm 1.1$                   & $67.67\pm 0.43$\\
$\Omega_{\rm m}$     & $0.3039\pm 0.0061$        & $0.3091^{+0.0091}_{-0.010}$     & $0.3110\pm 0.0057$\\
\hline
$E_{11 }$            & $-0.05^{+0.49}_{-0.56}$   & $0.05^{+0.29}_{-0.35}$          & $---$\\
$E_{22 }$            & $-1.38^{+2.5}_{-0.90}$    & $-0.45^{+1.2}_{-0.69}$          & $---$\\
$E_{21 }$            & $1.20^{+0.77}_{-2.1}$     & $0.43^{+0.60}_{-1.1}$           & $---$\\
$E_{12 }$            & $-0.13^{+0.53}_{-0.29}$   & $-0.13^{+0.43}_{-0.33}$         & $---$\\
\hline
\end{tabular}
\caption{Mean values and $68\%$ confidence level limits for cosmological and MG parameters obtained fitting the Planck early parameterizations to the combination of CMB and BAO data, assuming a $\Lambda$CDM background and including the effect of departures from GR in the expansion history. The last column shows the results obtained fitting the same data combination in a $\Lambda$CDM cosmology.}
\label{tab:MGbkg_pkearly}
\end{table}

\subsubsection{z\_flex parameterization}

\begin{figure}[!t]
	\centering
	\includegraphics[width=0.75\columnwidth]{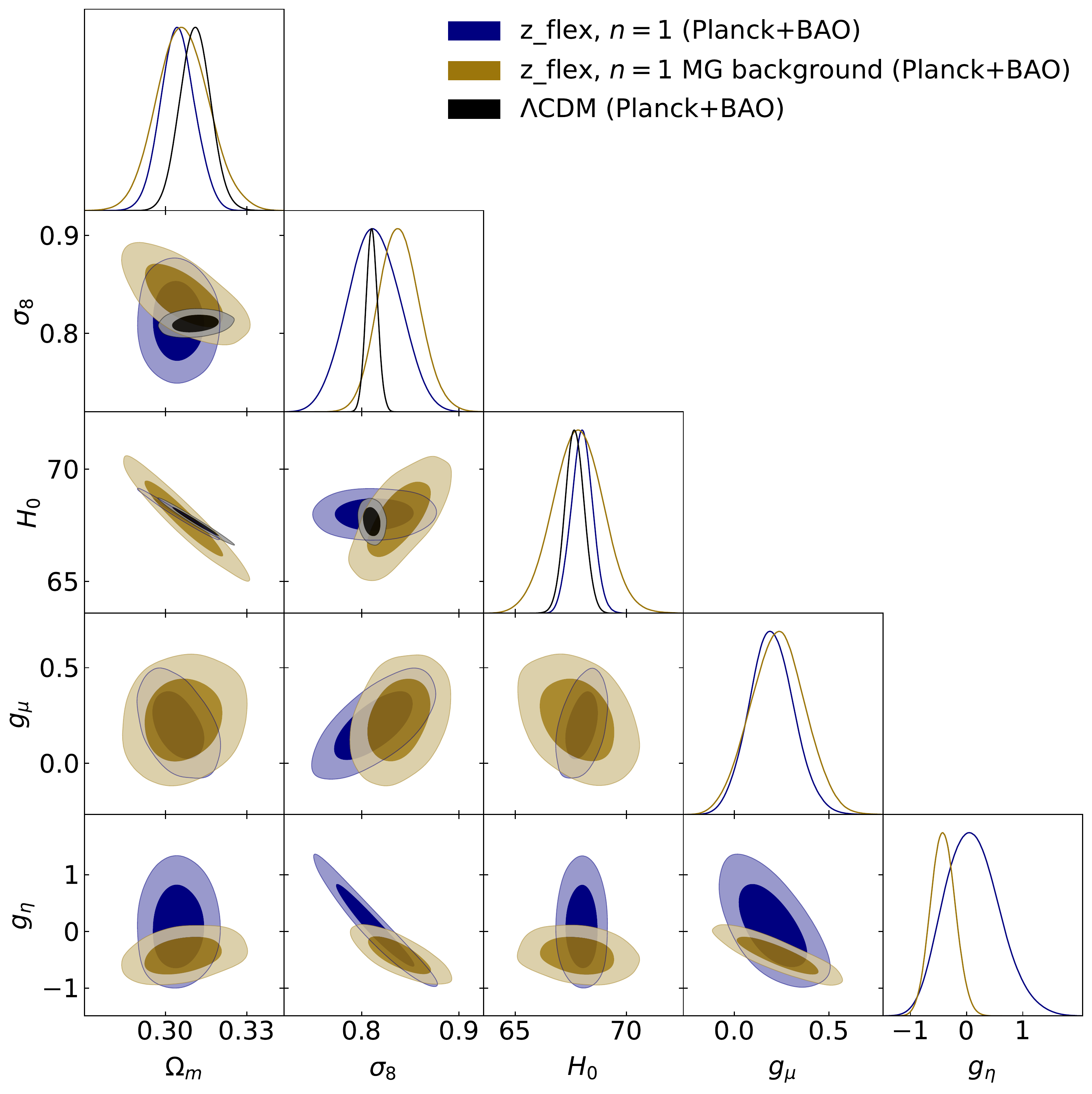}
	\begin{tabular}{cc}
	\includegraphics[width=0.45\columnwidth]{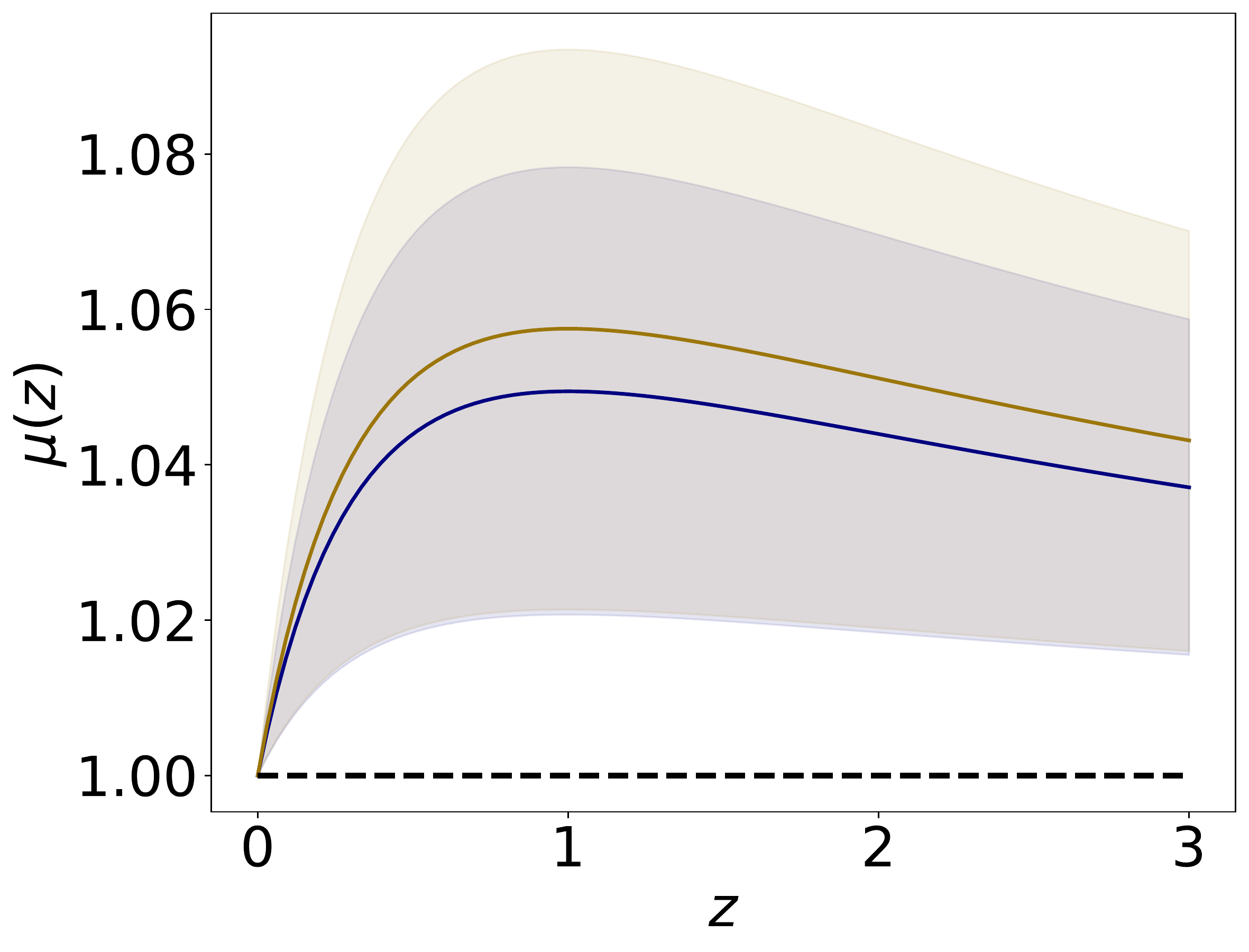} &
	\includegraphics[width=0.45\columnwidth]{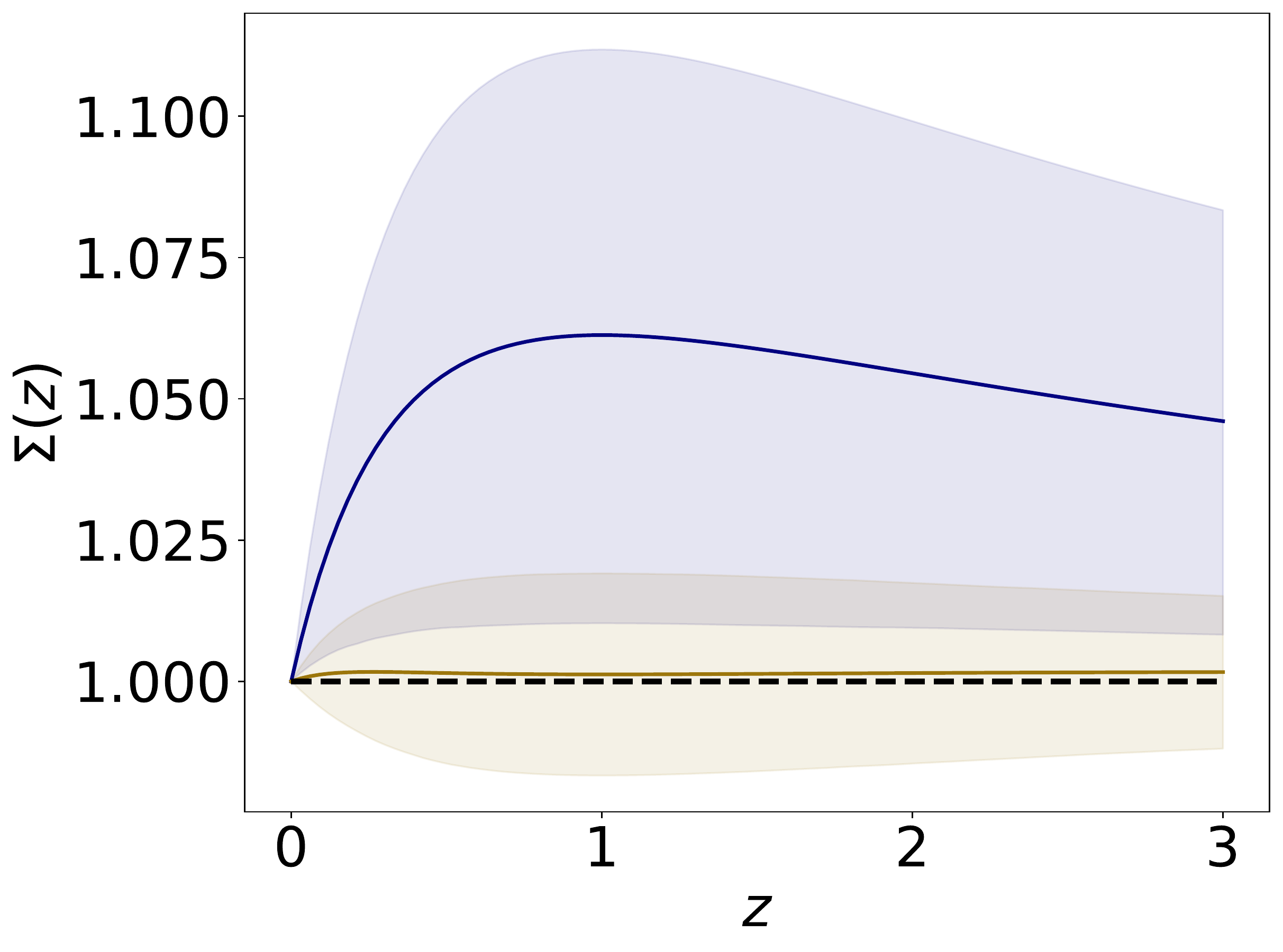}
	\end{tabular}
	\caption{Constraints on z\_flex when its effect on the background expansion are considered (brown), compared with the case with a $\Lambda$CDM background (blue) and with the $\Lambda$CDM limit (black). The bottom panel shows the constraints on the MG functions $\mu(z)$ and $\Sigma(z)$, while the top panel show the results obtained on the standard cosmological parameters. The data combination used here include CMB and BAO. 
	}\label{fig:MGbkg_zflex}
\end{figure}

We now explore the effect of including modifications to the background expansion when  constraining the z\_flex parameterization with $n=1$. The results are shown in \autoref{tab:MGbkg_zflex} and \autoref{fig:MGbkg_zflex}; we can notice how in this case the inclusion of the MG effects on the background leads to a $\Sigma(z)$ that has almost no deviation from the GR limit, while $\mu(z)$ is allowed to deviate more from its GR value, even though it is still compatible with it at $\approx2\sigma$. This shows that the background effect requires little or no deviation in the $\Sigma$ function, due to the interplay between the parameters of this case when entering the Friedmann equation, while the CMB data still require a high lensing effect and therefore cause $\mu(z)$ to increase its value.

The standard cosmological parameters have instead a behaviour similar to the Planck early case, with $\sigma_8$ that has a tighter bound when MG effects are included in the background, while $H_0$ bounds are relaxed due to new degeneracies with the MG parameters.

\begin{table}[]
    \centering
\begin{tabular} { l | c c | c}
\noalign{\vskip 3pt}\hline\hline
 & \multicolumn{2}{c|}{z\_flex} & $\Lambda$CDM\\
\hline
                     & $\Lambda$ background &  MG background    &  \\
\cline{2-4}

 Parameter           &  68\% limits              &  68\% limits                 &  68\% limits\\
\hline
$\Omega_{\rm c} h^2$ & $0.1184\pm 0.0010$        & $0.1185\pm 0.0015$        & $0.11933\pm 0.00093$\\
$\Omega_{\rm b} h^2$ & $0.02243\pm 0.00014$      & $0.02242\pm 0.00016$      & $0.02241\pm 0.00014$\\
$\ln10^{10}A_{\rm s}$& $3.028\pm 0.017$          & $3.034\pm 0.016$          & $3.047\pm 0.015$\\
$n_{\rm s}$          & $0.9670\pm 0.0040$        & $0.9668\pm 0.0049$        & $0.9663\pm 0.0037$\\
$\tau_{\rm reio}$    & $0.0497\pm 0.0080$        & $0.0522\pm 0.0078$        & $0.0561\pm 0.0074$\\
$\sigma_8$           & $0.813\pm 0.026$          & $0.838^{+0.020}_{-0.022}$ & $0.8103\pm 0.0060$\\
$H_0$                & $68.00\pm 0.48$           & $67.8\pm 1.1$             & $67.67\pm 0.43$\\
$\Omega_{\rm m}$     & $0.3046\pm 0.0063$        & $0.3065\pm 0.0096$        & $0.3110\pm 0.0057$\\
\hline
$g_\mu$              & $0.20\pm 0.12$            & $0.23\pm 0.14$            & $---$\\
$g_\eta$             & $0.098^{+0.44}_{-0.52}$   & $-0.42\pm 0.22$           & $---$\\
\hline
\end{tabular}
\caption{Mean values and $68\%$ confidence level limits for cosmological and MG parameters obtained fitting the z\_flex ($n=1$) parameterizations to the combination of CMB and BAO data, assuming a $\Lambda$CDM background and including the effect of departures from GR in the expansion history. The last column shows the results obtained fitting the same data combination in a $\Lambda$CDM cosmology.}
\label{tab:MGbkg_zflex}
\end{table}

\section{Conclusions}\label{sec:conclusions}
In this paper, we investigated the constraints achievable with CMB data from Planck on deviations from GR using two widely used parameterizations (Planck early and Planck late) and a new one (z\_flex). The latter ensures that the GR limit of the functions encoding deviations from the standard paradigm ($\mu=\eta=\Sigma=1$) is reached both at present time ($z=0$) and in the past ($z>>1$) where observations agree with Einstein's theory.

In order to obtain theoretical predictions to be compared with observational data, we implemented these parameterizations in a new code we developed, \texttt{MGCLASS II}, a modification of the publicly available \texttt{CLASS} code.

Using this code in combination with the public parameter estimation tools \texttt{Cobaya} and \texttt{MontePython}, we compared the theoretical predictions with the CMB data from Planck. We found that the Planck late and z\_flex parameterizations exhibit very similar behaviours, both in the recovered distributions for cosmological parameters and in the reconstructed trends for the $\mu(z)$ and $\Sigma(z)$ functions, while Planck early provides different trends for such functions at high redshift and looser bounds on cosmological parameters due to the higher freedom allowed by this parameterization.
This analysis also allows us to validate our implementation with the one done in \texttt{MGCAMB} which was used by the Planck Collaboration to obtain their results \cite{Ade:2015rim,Aghanim:2018eyx}, a comparison that we performed in \autoref{sec:validation}.

The \texttt{MGCLASS II} code allows to force the modifications on the $\Sigma(z)$ function, controlling the departure from the standard lensing effect, to vanish, a feature that is present in some classes of modified gravity models \cite{Brando:2019xbv}. Imposing such a constrain in our analysis, we found that the departure from the GR limit for the only free function $\mu(z)$ is increased. This is due to the fact that CMB data from Planck prefer an higher lensing with respect to what is predicticted in the standard paradigm; as this is not accounted anymore by modifications of $\Sigma(z)$, the increase in $\mu(z)$ modifies the density perturbation evolution in order to account for the higher lensing amplitude. Despite obtaining larger departures from the GR limit (with Planck late reaching $\approx2\sigma$), such a scenario would be disfavoured if low redshift observables such as weak lensing are included in the analysis, as the increase in $\mu$ generates a higher value of $\sigma_8$ and would therefore worsen the existing tension on this parameter between CMB and low redshift data.

Another feature we included in \texttt{MGCLASS II} is the possibility to account for the effect of departures from GR in the background expansion of the Universe. We have analysed the Planck early and z\_flex parameterizations in this context, highlighting how this affects the constraints on the MG functions, while also changing the bounds on standard cosmological parameters. In particular, we found that when MG effects are included in the background, $H_0$ has looser constraints with respect to the case where the background is assumed to $\Lambda$CDM, due to the new degeneracies with the MG parameters that are introduced.

In conclusion, this analysis shows that enforcing the MG parameterization to return to GR when expected (i.e. at present and very early times) does not significantly affect the constraints and the phenomenology of the deviations from GR. On the contrary, accounting for the effect of MG on the background expansion might significantly change constraints both on MG functions and cosmological parameter, and therefore such effects should be taken into account when possible in order to obtain realistic constraints on departures from GR.

In addition to these conclusions, the outcome of this work is a new code, \texttt{MGCLASS II}, able to account for parameterized deviations from GR and for specific MG models. It implements several MG models and parameterization not previously available in other public codes, thus allowing for these to be tested at the cosmological level. Moreovere, \texttt{MGCLASS II} follows a complementary approach with respect to existing codes. thus it can be used alongside these to cross-check the results obtained when analysing data and, therefore, to obtain more robust constraints on this kind of models.
\appendix

\section{Additional variants for some models}\label{sec:appendix}

Here we further elaborate and describe some variants of existing models or parameterisations that we implemented in this version of \texttt{MGCLASS} for the first time and were not introduced by other existing studies.

\subsection{Growth index with a $w\neq-1$}

The growth rate $ f=d\ln \rm D / d\ln  a$,  where $D$ is the growth of perturbations $\delta(z) = \delta_0  D(z)$, can be approximated by $ f=\Omega_{m}^\gamma(z)$ with $\Omega_{ m}(z)$ the matter density at $z$ and $\gamma$ named the growth index. It was introduced in \citep{1980lssu.book.....P} and found to well approximate the growth in $\Lambda$CDM when it is set to $\sim 0.545$ \citep{1990ApJS...74..831L,1991MNRAS.251..128L}. It takes different values in other modified gravity models \citep{2005PhRvD..72d3529L} and could be used as a trigger test for deviations from GR when comparing with observational data.

The implementation of the growth index $\gamma$ here follows the method described in \cite{Pogosian:2010tj}, where however a $\Lambda$CDM background is assumed. In \texttt{MGCLASS} we further allow a $w\neq-1$ following the equations below.

We start from the system of equations for scale independent modified gravity 

\begin{align}
 -k^2\,\Psi(a) &= \frac{4\pi\,G}{c^4}\,a^2\,\bar\rho(a)\,\Delta(a,\vec{k})\times\mu(a), \label{eq:pot_mu}\\ 
 \Phi(a) &= \Psi(a)\times\eta(a),  
  \label{eq:pot_eta}
\end{align}
 and the definition of the $\gamma$ parameterization for the growth, 
 
 \begin{equation}\label{eq:fgrowth}
 \Omega_m(a)^\gamma \, = \frac{d \log D_+}{d \log a}  ,
\end{equation}
where $D_{+} \equiv \Delta(a)_m/a$ is the growth rate, defined in terms of the matter density perturbation $\delta_m$.\\ 
 
Using then the second order equation for scale independent growth when $\eta(a)$=1

\begin{equation}
\Delta''(a)+\left[2+{H'(a) \over H(a)} \right]\Delta'(a)-{3 \over 2}{E_m(a) \over E(a)} \mu(a) \Delta(a)=0 \ ,
\label{equ:growthmu}
\end{equation}
 one can solve the equation in order to obtain $\Delta(a)$ and combine with \autoref{eq:pot_mu} to relate $\gamma$ to $\mu(a)$ as:

\begin{equation}
\mu(a)=\frac{2}{3}\Omega_m^{\gamma-1}(a)\left[\Omega_m^{\gamma}(a)+2+\frac{H'(a)}{H(a)}+\gamma\frac{\Omega_m'(a)}{\Omega_m(a)}+\gamma'\ln\left(\Omega_m(a)\right)\right]\, ,\label{eq:mugamma}
\end{equation}
where in the assumption of a flat Universe and constant $w$
\begin{align}
\Omega_m(a) &=\frac{H^2_0}{H^2}\Omega_m a^{-3}\,, \\    \frac{\Omega_m'(a)}{\Omega_m(a)} &=-3 -2\frac{H'(a)}{H(a)}\,,\\ \frac{H'(a)}{H(a)} &= -\frac{3}{2}\Omega_m(a)-\frac{3}{2}(1+w)\Omega_{DE}(a)\,,\\ 
\Omega_{DE}(a) &= \frac{H^2_0}{H^2}(1-\Omega_m a)^{-3(1+w)}\,.
\end{align}

Therefore the relation between $\mu(a,k)$ and $\gamma$ becomes

\begin{equation}
\mu(a)=\frac{2}{3}\Omega_m^{\gamma-1}(a)\left[\Omega_m^{\gamma}(a)+2 -3\gamma + 3(\gamma - \frac{1}{2})\left(\Omega_m(a)+(1+w)\Omega_{DE})\right)\right]
\end{equation}

\begin{figure}[!h]
	\centering
	\includegraphics[width=0.5\columnwidth]{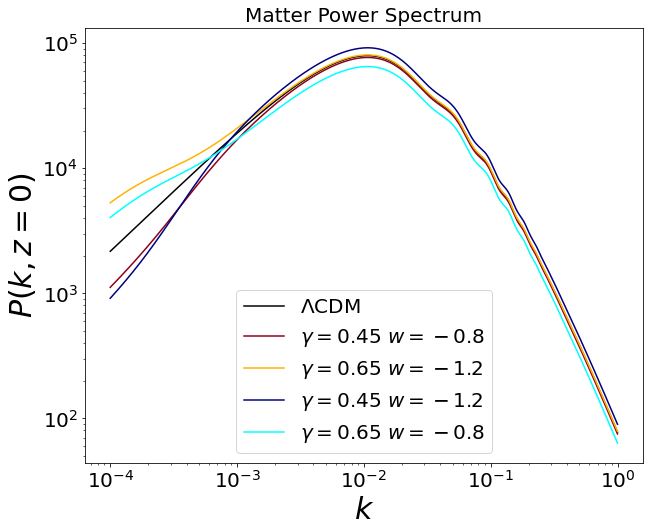} 
	\caption{The matter power spectrum for different values of the growth index $\gamma$ along with different values of the dark energy EoS parameter $w$ in comparison with $\Lambda$CDM power spectrum. The cosmological parameters are fixed to the mean values of Planck 2018 \cite{Aghanim:2018eyx} }\label{fig:gam_pk} 
\end{figure}

We show in \autoref{fig:gam_pk} the power spectrum for different combinations of the growth index $\gamma$ and the dark energy EoS parameter $w$.

\subsection{Designer approach models}\label{dam}

Below we describe the methods implemented to account for the influence of some of the specific models on the evolution of the background of the Universe. For the models described here, the general method shown in \autoref{sec:MGgeneralities} could not be applied with the same accuracy.

\subsubsection{The JBD model}

The JBD cosmological model was formulated as the first scalar-tensor theory of gravity \citep{Brans:1961sx}. In principle, it only features one more degree of freedom with respect to GR, the JBD parameter $\omega_{\rm JBD}$, which is constant both in space and time.

 Its Lagrangian reads
\begin{equation}
{ \mathcal{L}_{\rm JBD}} = \frac{1}{16\pi G}\left({  \phi}R - \frac{\omega_{\rm JBD}}{  \phi} \partial_\mu {  \phi}\partial^\mu {  \phi}\right) - V({  \phi}) + { \mathcal{L}}_{\rm fluid}\,.
\end{equation}

GR is a particular case of the JBD theory, corresponding to $\omega_{\rm JBD}  = \infty$. Also, usually the potential ($V$) is set to be constant, mimicking a $\Lambda$ term that drives the late-time acceleration.

The equations of motion are in this case :
\begin{equation}\label{Hjbd}
{H}^2 + {H} \frac{\dot{  \phi}}{  \phi} = \frac{\omega_{\rm JBD}}{6}\left(\frac{\dot{  \phi}}{  \phi}\right)^2 + \frac{8 \pi G}{3}\frac{\rho}{  \phi}\, , 
\end{equation}
\begin{equation}\label{phimot}
\ddot{ \phi} + 3 {H} \dot{  \phi} = \frac{8 \pi G}{2 \omega_{\rm JBD}+3}(\rho - 3 p)\, ,
\end{equation}
with $V$ and matter fluids absorbed in $\rho$ and $P$. The field $\phi$ needs to meet conditions given by Solar System constraints which translate into an initial value
 \begin{equation}
\phi_0 = \frac{2\omega_{\rm JBD}+4}{2\omega_{\rm JBD}+3}\,.
\end{equation}

By mapping this theory into the $\mu$ and $\eta$ function, we find them to be only dependent on the redshift and the two functions read \citep{Boisseau:2000pr}

\begin{equation}
\mu (z)= \frac{2\omega_{\rm JBD}+4}{2\omega_{\rm JBD}+3}\frac{1}{\phi(z)}\,,
\end{equation}

\begin{equation}
\eta(z) = \frac{\omega_{\rm JBD}+1}{\omega_{\rm JBD}+2}\,.
\end{equation}

To implement the background evolution within this framework we followed the designer approach method \citep{Lima:2015xia}
in which, simultaneously neglecting the $\phi^{\prime \prime}$ and $\phi^{\prime 2}$ terms, \autoref{phimot} is expressed as

\begin{equation}{\label{fieldfrwapprox1}}
 \frac{\phi^{\prime}}{\phi}\left({1 - \frac{1}{2}\frac{\Omega_{\rm{m}}a^{-3}}{1 - \Omega_{\rm{m}} + a^{-3}\Omega_{\rm{m}}}}\right) = \frac{4({1-\Omega_{\rm{m}}}+a^{-3}\Omega_{\rm{m}})}{d({1 - \Omega_{\rm{m}}}+a^{-3}\Omega_{\rm{m}})},
\end{equation}
where $d = 2 \omega_{\rm BD} + 3$.

The solution for the scalar field will be a fully analytical expression, given by
\begin{equation}{\label{phisolsimpleminus1}}
 \phi(a) = \phi(a_i) g(a_i)^{-1} g(a),
\end{equation}
where $\phi(a_i)$ is the scalar field value at a high redshift $z_i$ set to be $\phi_{0}$ at $a_i = 1$, while the function $g(a)$ is given by 
\begin{equation}
g(a) = a^{\frac{2}{d}}({2a^{3}({1-\Omega_{\rm{m}}}+\Omega_{\rm{m}}})^{\frac{2}{3d}}\, ,
\end{equation}
and the latter is used to get $\phi$ in order to solve for $H$ in \autoref{Hjbd}.

\subsubsection{The nDGP model}

The Dvali-Gabadadze-Porrati (DGP) is a model in which gravity leaks off the 4-dimensional Minkowski brane where it behaves like GR into the 5-dimensional “bulk” Minkowski space-time at large scales governed by a crossover scale $r_c$ \citep{Dvali:2000hr}. 

The model is described by the action
\begin{equation}
S = M_{\ast}^3\int d^5\sqrt{-\gamma} \, \mathcal{R} + \int
d^4x\sqrt{-g}\left(\mathcal{E}_4 + M_{P}^{2}R +
\mathcal{L}_{SM}\right),
\end{equation}
where $M_{\ast}^3$ is the Planck mass defining a certain $5$-dimensional scale, $\mathcal{R}$ the
equivalent Ricci scalar of the higher dimensional theory and $\gamma_{ab}$ the
$5$-dimensional metric, while $M_{P}$, $R$, $g$ and $\mathcal{E}_4 = M_{Pl}^2\Lambda$ terms are the ones of the induced 4D-Einstein-Hilbert action and $\mathcal{L}_{SM}$ the matter action part. For small values of the threshold $r_c\sim M_{P}^2/M_{\ast}^3$, classical GR is recovered while for larger values $r_c$ becomes of comparable
magnitude with the one in the fifth dimension at very large distances modifying GR.

As a result, while the conservation equation remains the same as GR, the equation of motion is modified, leading to a modified Friedmann equation \cite{Deffayet:2000uy}

\begin{equation}\label{Hndgp}
H^2 = \frac{8\pi G}{3}\rho_{tot} + \epsilon \frac{H}{r_c} + \frac{\Lambda}{3}\,.
\end{equation}

with the $\epsilon$ taking $\pm1$ values. We adopt the solution for $\epsilon = -1$, called the normal branch. In this branch, acceleration is achieved through a cosmological constant as in GR.

This results, in the linear quasistatic limit, in the possibility to write the functions $\mu$ and $\eta$ as \citep{Koyama:2005kd} :

\begin{equation}
\mu =   \left(1 + \frac{1}{3 \beta} \right)\,, 
\end{equation}

\begin{equation}
\eta =  \frac{\left(1 - \frac{1}{3 \beta} \right)}{ \left(1 + \frac{1}{3 \beta} \right)}\,,
\end{equation}
where  
\begin{equation}
\beta = 1 - 2 H r_c \left( 1 + \frac{\dot{H}}{3 H^2} \right)\,,
\end{equation}
while to implement the background evolution for this model, we solved the \autoref{Hndgp} for $H$ which now becomes

\begin{equation}\label{Fndgp}
H = \sqrt{\rho_{tot}-\frac{1}{4r^2_{c}}} -\sqrt{\frac{1}{4r^2_{c}}}\,,
\end{equation}
and we used the fact that the conservation equation stays the same as GR to derive $\dot{H}$ to close the system.

We finally use the modified Friedmann equation of \autoref{Fndgp} at $z = 0$ to get $\Omega_{\Lambda,0}$

\subsubsection{The K-mouflage model}

K-mouflage theories are built complementing simple K-essence
scenarios with a universal coupling of the scalar field $\phi$ to matter \citep{Babichev:2009ee}; they are defined by the action
\begin{align}
S &= \int d^4 x \sqrt{-{g}} \left[ \frac{{M}_{\rm Pl}^2}{2} {R} + {\cal M}^4 K({\chi}) \right]  +S_{\rm m}(\psi_i, g_{\mu\nu})\, ,
\label{eq:actkm}
\end{align}
where ${M}_{\rm Pl}$ is the Planck mass, ${\cal M}^4$ is the energy scale of the scalar field, ${g}_{\mu \nu}$ is the Jordan frame metric, $\tilde{g}_{\mu \nu}$ is the Einstein frame metric with $\tilde{g}_{\mu \nu}=A^2(\phi){g}_{\mu \nu}$, $S_{m}$ is the action of the matter fields $\psi_{m}^{(i)}$, ${\chi}$ is defined as
\begin{equation}
{\chi} = - \frac{{g}^{\mu\nu} \partial_{\mu}\phi\partial_{\nu}\phi}{2 {\cal M}^4} \ ,
\end{equation}
and ${\cal M}^4 K$ is the non-standard kinetic term of the scalar field. \\
The amount of deviation from $\Lambda$CDM at the background level and at linear order in perturbation theory can be expressed in terms of two time-dependent functions \citep{Brax:2014yla}
\begin{equation}
\label{epsilon2-def}
\epsilon_2 = \frac{\ln {A}}{\ln a} \ , \ \ \epsilon_1 = \frac{2}{{K}'} \left(\epsilon_2  {M}_{\rm Pl} \left(\frac{{\phi}}{\ln a}\right)^{-1} \right)^2 \ , 
\end{equation}
where a prime indicates derivatives with respect to ${\chi}$. 
The K-mouflage Friedmann equations therefore read
\begin{eqnarray}
3 {M}_{\rm Pl}^2 H^2 = \frac{{A}^2}{(1-\epsilon_2)^2} \;
\left({\rho}_m +\frac{{\cal M}^4}{{A}^4} \left(2 {\chi} {K'} - {K}\right)\right) \, ,  \\
-2 {M}_{\rm Pl}^2 \dot{H} = \frac{{A}^2}{(1-\epsilon_2)^2} \;
\left({\rho}_m +\frac{{\cal M}^4}{{A}^4} \left(2 {\chi} {K'} - {K}\right) + {\cal M}^4 \, K \right) \,,
\label{E00}
\end{eqnarray}
and the MG functions $\mu$ and $\eta$ become \citep{2019JCAP...05..027B}
\begin{equation}
\label{mu_Sigma_KM}
\mu(a) = (1+ \epsilon_1) {A}^2 \, , \ \ \ \eta(a)=  \frac{(1- \epsilon_1)}{(1+ \epsilon_1)} {A}^2 \, . 
\end{equation}

In our implementation we choose one of the parameterisation proposed in \citep{Brax:2014wla} for $K(\chi)$ along with the coupling $A(\phi)$:

\begin{equation}
K(\chi) = -1 + \chi + K_0 \, \chi^m \, , \ \ \ 
A(\phi)= 1 + \beta \phi\,,
\end{equation}
with $K_0$ and $\beta$ as free parameters while m was fixed in our code to 2 and its derivative $K' = 1 + K_0 \, \chi^m$.\\

This background parameterization shows in the value of the associated effective energy density and pressure term necessary for us to implement the evolution of the background, with ${\cal M}^4$ of the order of the current energy density $\rho_\Lambda$, to retrieve the late-time accelerated expansion of the Universe

\begin{equation}
\rho_\phi = \frac{\rho_\Lambda}{(1+\beta a)^4}\left( -1 + \chi + K_0 \chi^m +(1+\beta a)^2 \dot{\phi}^2 (1 + m K_0\phi^{m-1})\right) \, ,
\end{equation}
\begin{equation}
p_{\phi}= K (\chi) \frac{\rho_{\Lambda}}{(1+\beta a)^4}\,,
\end{equation}
and that of  the two time-dependent functions $\epsilon_1$ and $\epsilon_2$ used to construct the MG functions

\begin{equation}
\epsilon_1 = \frac{2 \beta^2}{1 + m K_0 \chi^{m-1}} \, , \ \ \ 
\epsilon_2 = \frac{a \beta}{1+\beta a}
\end{equation}

\begin{figure}[!h]
	\centering
	\begin{tabular}{cc}
	\includegraphics[width=0.45\columnwidth]{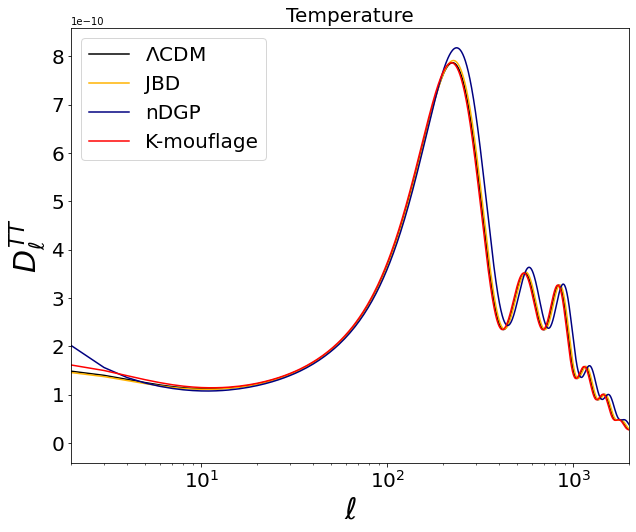} &
	\includegraphics[width=0.45\columnwidth]{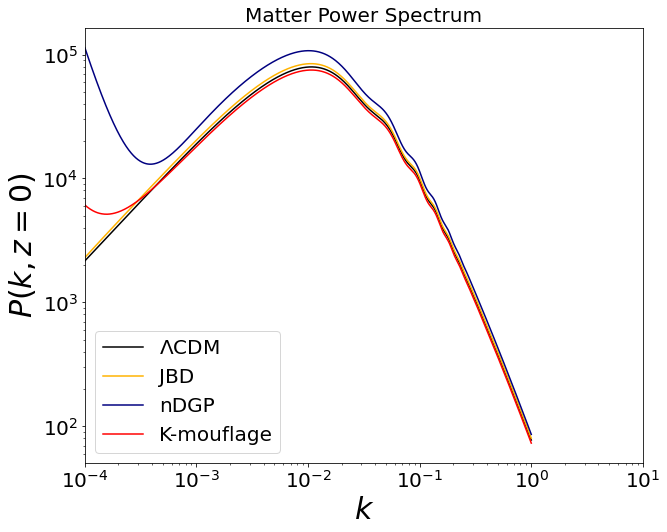} 
	\end{tabular}
	\includegraphics[width=0.45\columnwidth]{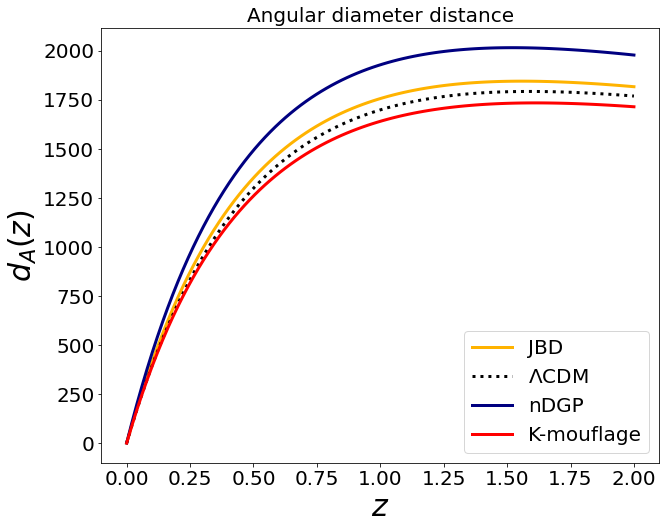}
	\caption{Comparison of three specific models, nDGP, JBD and K-mouflage, available in \texttt{MGCLASS II}. The top left panel shows the temperature power spectra, the top right panel shows the matter power spectrum, while the bottom panel shows the Hubble parameter function. For this plot, the cosmological parameters are fixed to the mean values of Planck 2018 \cite{Aghanim:2018eyx}, while the nDGP are set to be $\{r_c\}=\{10000.0\}$, the JBD are set to be $\{w_{JBD}\}=\{75.0\}$ and the K-mouflage are set to be $\{\beta,K_0\}=\{0.05,0.1\}$.
	}\label{fig:models_checks} \end{figure}

In \autoref{fig:models_checks} we show the angular CMB temperature power spectrum, the matter power spectrum and the Hubble parameter evolution for this model, alongside the JBD and nDGP ones, in comparison with $\Lambda$CDM.

\

\subsection{The z\_xpans parameterization}

The z\_xpans is a parameterization of the modified gravity potentials $\mu$ and $\eta$ inspired from the Taylor expansion approach where the MG functions are expanded as a function of redshift up to a given order. We implement a z\_xpans early case, basically extending the Planck early parameterizion to second order, and a late case adding an extra term to the Planck late parameterization. Notice that the latter is only loosely inspired to a Taylor expansion, as this is not really the way of expanding around a specific value of $\Omega_{\rm DE}$. For both cases, we have four parameters ($T_1$, $T_2$, $T_3$ and $T_4$), which allows high flexibility.

We show here the MG functions $\mu$ and $\eta$ and their derivatives in the late case:

\begin{equation}
\mu(a)  = 1+T_1\Omega_{\Lambda}^n + T_2\Omega_{\Lambda}^{2n}\,, \label{muansatz}
\end{equation}

\begin{equation}
\eta(a)  = 1+T_3\Omega_{\Lambda}^n + T_4\Omega_{\Lambda}^{2n}\,, \label{etaansatz}
\end{equation}

\begin{equation}
\dot{\mu}(a)  = n\,\dot{\Omega}_{\Lambda}\,T_1\Omega_{\Lambda}^{n-1} + 2n\,\dot{\Omega}_{\Lambda}\,T_2\Omega_{\Lambda}^{2n-1}\,, \label{muansatzdot}
\end{equation}

\begin{equation}
\dot{\eta}(a)  = n\,\dot{\Omega}_{\Lambda}\,T_3\Omega_{\Lambda}^{n-1} + 2n\,\dot{\Omega}_{\Lambda}\,T_4\Omega_{\Lambda}^{2n-1}\,.  \label{etaansatzdot}
\end{equation}\\\\

\begin{figure}[!h]
	\centering
	\begin{tabular}{cc}
	\includegraphics[width=0.45\columnwidth]{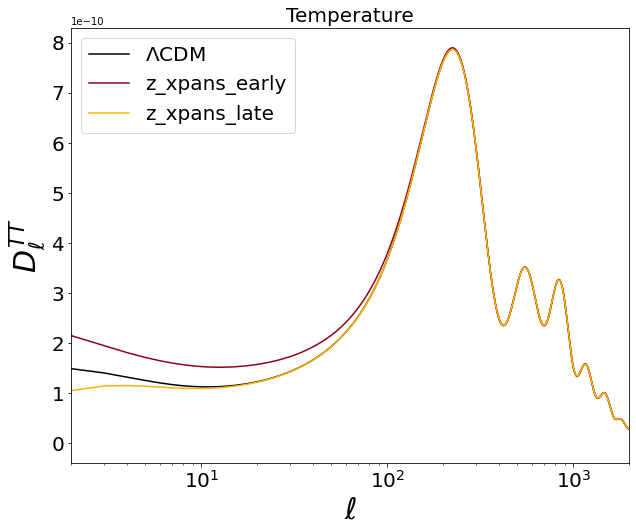} &
	\includegraphics[width=0.45\columnwidth]{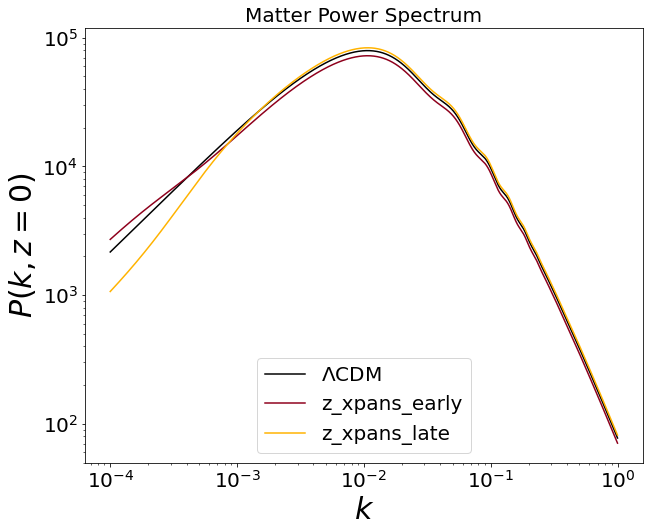}
	\end{tabular}
	\caption{Comparison of the two different phenomenological assumptions available in \texttt{MGCLASS II} for the z\_xpans parameterization. The left panel shows the temperature power spectra, while the right panel shows the matter power spectra. For this plot, the cosmological parameters are fixed to the mean values of Planck 2018 \cite{Aghanim:2018eyx}, while the z\_xpans are set to be $\{T_1,T_2,T_3,T_4\}=\{0.3,0.1,0.3,0.1\}$. }\label{fig:zxpans_clmpk} \end{figure}

Below we show in \autoref{fig:zxpans_clmpk} the angular CMB temperature power spectrum and the matter power spectrum for the two different z\_xpans parameterisations in comparison with $\Lambda$CDM.

\

\acknowledgments

MM has received the support from the fellowship “la Caixa” Foundation (ID 100010434), with fellowship code LCF/BQ/PI19/11690015.

\bibliographystyle{unsrt}
\bibliography{references}

\end{document}